%% file: main.tex
\RequirePackage[hyphens]{url}
\RequirePackage[sort,nocompress]{cite}

\documentclass[hyperref={bookmarks=true,breaklinks=true,letterpaper=true,colorlinks,citecolor=blue,linkcolor=blue,urlcolor=blue},sigconf]{acmart}

\usepackage[misc]{ifsym}
\usepackage{caption}
\usepackage{fancyhdr}
\usepackage{cite}
\usepackage{amsmath,amssymb,amsfonts}
\usepackage{algorithmic}
\usepackage{graphicx}
\usepackage{textcomp}
\usepackage{xcolor}
\usepackage{CJKutf8}
\usepackage{array}
\usepackage{booktabs}
\usepackage{rotating}
\usepackage{tablefootnote}
\usepackage{multirow}
\usepackage{ragged2e}
\usepackage{color}
\usepackage{makecell}
\usepackage{vcell}
\usepackage{diagbox}
\usepackage{pifont}
\usepackage{tabularray}
\usepackage[lined,boxed,commentsnumbered,ruled]{algorithm2e}
\usepackage{bbding}

\input{signature}

\begin{document}

\input{title}

\input{MICRO/1_Introduction}

\input{MICRO/2_Background}

\input{MICRO/3_Motivation}

\input{MICRO/4_Innovation_Architecture.tex}

\input{MICRO/6_Implementation}
\input{MICRO/7_Evaluation_Methodology}

\input{MICRO/8_Evaluation}
\input{MICRO/9_Related_Work}
\input{MICRO/10_Conclusion}

\input{MICRO/11_Acknowledgments}

\bibliographystyle{ACM-Reference-Format}
\bibliography{MICRO/refs}

\end{document}

%% file: signature.tex
\AtBeginDocument{%
  }

\copyrightyear{2023} 

\acmYear{2023}
\setcopyright{rightsretained} 
\acmConference[MICRO '23]{56th Annual IEEE/ACM International Symposium on
Microarchitecture}{October 28-November 1, 2023}{Toronto, ON, Canada}
\acmBooktitle{56th Annual IEEE/ACM International Symposium on
Microarchitecture (MICRO '23), October 28-November 1, 2023, Toronto, ON,
Canada}
\acmDOI{10.1145/3613424.3614246}
\acmISBN{979-8-4007-0329-4/23/10}

%% file: title.tex
% \affil{1. School of Integrated Circuits, Tsinghua University, Beijing, China, 100084}
% \affil{$^{\textrm{\Letter}}$ Corresponding  Author, hu\_yang@tsinghua.edu.cn}

\title{Towards Efficient Control Flow Handling in Spatial Architecture via Architecting the Control Flow Plane}

\author{
    Jinyi Deng\textsuperscript{1},
    Xinru Tang\textsuperscript{1},
    Jiahao Zhang\textsuperscript{1},
    Yuxuan Li\textsuperscript{1},
    Linyun Zhang\textsuperscript{1},
    Boxiao Han\textsuperscript{3},
    Hongjun He\textsuperscript{3},
    Fengbin Tu\textsuperscript{2},
    Leibo Liu\textsuperscript{1},
    Shaojun Wei\textsuperscript{1},
    Yang Hu\textsuperscript{1\textrm{\Letter}},
    and Shouyi Yin\textsuperscript{1,4}
}

\affiliation{%
    \institution{\textsuperscript{1}School of Integrated Circuits, Tsinghua University, Beijing, China}
    \country{}
}
\affiliation{%
    \institution{\textsuperscript{2}Hong Kong University of Science and Technology, Hongkong, China}
    \country{}
}
\affiliation{%
    \institution{\textsuperscript{3}China Mobile Research Institute, Beijing, China}
    \country{}
}
\affiliation{%
    \institution{\textsuperscript{4}Shanghai AI Lab, Shanghai, China}
    \country{}
}
\affiliation{
    \institution{
        \textsuperscript{ \textrm{\Letter} } Corresponding  Author, hu\_yang@tsinghua.edu.cn}
    \country{}
}

\renewcommand{\shortauthors}{Jinyi Deng et al.}
%作者序列的简称

\begin{abstract}

Spatial architecture is a high-performance architecture that uses control flow graphs and data flow graphs as the computational model and producer/consumer models as the execution models. However, existing spatial architectures suffer from control flow handling challenges. Upon categorizing their PE execution models, we find that they lack autonomous, peer-to-peer, and temporally loosely-coupled control flow handling capability. This leads to limited performance in intensive control programs.

A spatial architecture, Marionette, is proposed, with an explicit-designed control flow plane. The Control Flow Plane enables autonomous, peer-to-peer and temporally loosely-coupled control flow handling. The Proactive PE Configuration ensures computation-overlapped and timely configuration to improve handling Branch Divergence. The Agile PE Assignment enhance the pipeline performance of Imperfect Loops. We develop full stack of Marionette (ISA, compiler, simulator, RTL) and demonstrate that in a variety of challenging intensive control programs, compared to state-of-the-art spatial architectures, Marionette outperforms Softbrain, TIA, REVEL, and RipTide by geomean 2.88×, 3.38×, 1.55×, and 2.66×.

\end{abstract}


\ccsdesc[500]{Computer systems organization~Reconfigurable computing}
\ccsdesc[300]{Theory of computation~Models of computation}

\keywords{spatial architecture, control flow, control plane, coarse-grained reconfigurable array}

\maketitle

%% file: MICRO/1_Introduction.tex
\section{introduction}\label{sec:1}

Moore's Law has propelled progress in conventional processors for numerous decades, and the formerly exponential growth is now diminishing. Fortunately, spatial architectures, such as Coarse-Grained Reconfigurable Array (CGRA) \cite{2021uecgre,2021fifer,Nguyene2020pipette,torng2020dac,gobieski2021snafu,bandara2022revamp,karunaratne2017hycube}, Reconfigurable Dataflow Processors \cite{2017plasticine,karunaratne20194d,TIA,2014sSGMF}, and Systolic Arrays \cite{2014FPCA,2017stream,2006tartan,piperench}, exhibit great promise owing to their inherent flexibility and remarkable performance. 

Spatial architectures represent a category of accelerators that harness the immense potential of high computational parallelism through direct intercommunication among an array of processing engines (PEs). The software programs are transformed into control flow graphs (CFGs) and data flow graphs (DFGs) to facilitate their compilation. The CFG captures the program's control dependencies, including loops and conditions. The widely prevalent execution model for spatial architectures is the producer/consumer pipeline model \cite{chen2016eyeriss}. The CFGs and DFGs are assigned to PEs and interconnect networks. This mapping relationship is expressed through carefully distributed instructions (configurations) among the PEs and interconnect networks. Spatial architecture can mitigate data transfers from memory, thus bypassing the storage bottleneck and facilitating the relentless advancement of computing power. Furthermore, its producer/consumer pipeline model excels in accelerating kernels with conventional data-level parallelism, bringing noteworthy performance benefits in many domains such as artificial intelligence (AI), mobile communication, and image processing \cite{bae2018auto,fan2018stream,gao2016hrl,pellauer2019buffets,vasilyev2016evaluating,nicol2017coarse}.

%They also can handle simple control flow such as non-nested loops well by statically mapping BBs to spatial PEs. Recent work has advocated that CGRA could be a next-generation general-purpose processor (GPP)\cite{2021uecgre,gobieski2022riptide}. 

However, contemporary applications impose heightened demands on the spatial architecture's capacity to adeptly handle control flow. On the one hand, modern applications across pivotal domains, such as mobile communication, computer vision, bioinformatics, and general-purpose kernels, exhibit intricate control flow patterns, involving branches or nested loops, as shown in Table \ref{table:tab1}. On the other hand, AI also has higher requirements for control flow processing capabilities. Tenstorrent has introduced a cutting-edge framework for big AI models, encompassing Dynamic Sparsity, Conditional Execution, and Dynamic Routing \cite{tenstorrent2022web}. FlexMoE \cite{nie2023flexmoe} has further proposed an innovative scheduling module that enables dynamic mapping of the model-to-hardware allocation based on real-time dataflow over the existing DNN runtime. 

\input{MICRO/Complex-CF}

Unfortunately, spatial architectures exhibit limitations in effectively handling control flows. To tackle this challenge, this work strives to comprehensively pinpoint the fundamental architectural limitations of the hardware execution model of spatial architecture, at both array-level and PE-level. We survey the representative spatial architectures in the past decade, categorizing their PE execution models into two distinct paradigms, namely von Neumann PE and dataflow PE. Furthermore, we conduct an in-depth analysis of these two models in the context of handling two typical control flow scenarios, namely Branch Divergence and Imperfect Loop.

Our observations indicate that conventional PE execution models cause significant PE idleness while handling above typical control flows. This limitation stems from the fact that: (1) von Neumann PEs lack the capability to autonomously initiate configuration changes in other PEs, and that there exists no direct channel for control information transfer between PEs; (2) The distinctive token utilized by dataflow PEs results in a temporally and spatially close-knit coupling of control flow and data flow, consequently constraining control flow transfer. We advocate that an ideal PE for control flow handling is expected to (1) \textit{\textbf{autonomously change the configurations of other PEs}}, (2) incorporate a \textit{\textbf{peer-to-peer control flow path}} to enable agile control information transmission, and (3) be \textit{\textbf{temporally loosely-coupled with dataflow path}}.

To achieve the above objectives, our insight is that the control flow handling in spatial architectures should stand out from the current hybrid design that mixes control flow handling and data flow handling. This calls for a new architectural scheme, which \textit{\textbf{decouples the control flow handling and dataflow handling}}. 
We introduce Marionette, a spatial architecture design with decoupled control flow plan and data flow plane. Specifically, we architect a control flow plane for existing spatial architecture, which includes redefined PE architecture and features. We believe a decoupled control flow plane is naturally beneficial for autonomous, peer-to-peer, and temporally loosely-coupled control flow handling capability.

\textbf{Control Flow Plane:} The control and configuration components of Marionette have been consolidated into a control flow plane, which completely decouples control flow handling from data flow handling. To enable autonomous, peer-to-peer and temporally loosely-coupled control flow handling, the control flow plane incorporates three distinct PE micro-architectures and a specialized control network.

\textbf{Proactive PE Configuration:} To computation-overlapped and timely configuration, we devise the Control Flow Sender, which timely transfer control flow to hasten the configuration of subsequent PEs. This allows for executing current-stage computation and next-stage configuration in the same stage. Consequently, Marionette achieves a high PE utilization rate in Branch Divergence.

\textbf{Agile PE Assignment:} Given the solid foundation for control flow handling capabilities provided by the Marionette control flow plane, we optimize the Marionette scheduling strategy and develope a Control Flow Scheduler. This enhancement renders Marionette highly flexible in constructing basic block pipelines, leading to a significant improvement in PE utilization when processing Imperfect Loops.

The contributions of this work are as follows:

\begin{enumerate}
%\item To the best of our knowledge, this is the first work that comprehensively rethinks the role of control flow in the spatial architecture execution model.
\item We present a taxonomy that categorizes PE execution models of spatial architectures based on prior research.
%\item We rephrase Amdahl's Law on CGRA from a control flow perspective and define the ``control flow plane" of CGRA. The implicit design of the control plane of the execution model leads to a terrible Amdahl's Law speedup.
\item We synopsize the control flow predicaments encountered by extant spatial architectures when confronted with demanding control flow applications, specifically, the paucity of autonomous, peer-to-peer, and temporally loosely-coupled control flow handling capability. To the best of our knowledge, this study represents the first comprehensive reexamination of the role of control flow in execution models for spatial architectures.
\item We enhance the execution model of spatial architectures by adopting a decoupled control flow plane and introducing three corresponding innovative features. Marionette, our software-defined hardware solution, realizes each features, and we implement a full stack including compiler, simulator, and RTL design.
\item We conduct a comprehensive evaluation that includes the acceleration effect of each innovation features. Compared to the state-of-the-art architectures, Marionette outperforms Softbrain, TIA, REVEL, and RipTide by geomean 2.88×, 3.38×, 1.55×, and 2.66× across a variety of intensive control flow applications.
\end{enumerate}

%The rest of this paper is organized as follows: Section \ref{sec:2} describes the computational model and execution model of CGRA. Section \ref{sec:3} analyzes the three challenges of the existing execution model plane in complex-CF applications. Section \ref{sec:4} gives an optimized control plane execution model and shows a framework for implementing software-defined hardware and the associated ISA. Then, we develop our Marionette in Section \ref{sec:5}. Finally, Sections \ref{sec:6}, \ref{sec:7}, \ref{sec:8} and \ref{sec:9} present the Implementation, methodology, evaluation, and other related work.

%% file: MICRO/Complex-CF.tex
\begin{table}\Large

    \setlength{\abovecaptionskip}{0pt}
    \renewcommand{\arraystretch}{1.15}
    \centering
    \caption{\label{table:tab1}Control flow forms across modern applications.}
    \resizebox{1\linewidth}{!}{%
    \begin{tabular}{l|ccc} 
    \toprule[2pt]
    \textbf{Domain}  & \textbf{Workload}  & \textbf{Intensive Branch}  & \textbf{Intensive Loop} \\ 
    \hline
    \multirow{5}{*}{\begin{tabular}[c]{@{}l@{}}\textbf{General}\vspace{-0.4em}\\\textbf{purpose}\end{tabular}}     & Merge Sort \cite{2014machsuite}  & {\begin{tabular}[c]{@{}c@{}}Nested branches\vspace{-0.4em}\\Innermost\end{tabular}}      & {\begin{tabular}[c]{@{}c@{}}Under branch\vspace{-0.4em} \\Imperfect nested\end{tabular}}      \\
                                                                                                    \cline{2-4}
                                                                                                    &\multirow{2}{*}{ FFT \cite{2014machsuite}}         & \multirow{2}{*}{Innermost}                 & \multirow{2}{*}{Imperfect nested}     \\ \\
                                                                                                    \cline{2-4}
                                                                                                    &\multirow{2}{*}{ Viterbi \cite{2014machsuite}}     &\multirow{2}{*}{Innermost }                    & \multirow{2}{*}{Imperfect nested}     \\ \\
                                                                                                    \cline{2-4}
                                                                                                    &\multirow{2}{*}{ GEMM \cite{2014machsuite}}     &\multirow{2}{*}{N/A }                    & \multirow{2}{*}{Imperfect nested}     \\ \\
    \hline
    \textbf{Bioinformatics}                                                      & NW \cite{2014machsuite}          & {\begin{tabular}[c]{@{}c@{}}Nested branches\vspace{-0.4em}\\Innermost\end{tabular}}     & Nested      \\
    \hline  
    {\begin{tabular}[c]{@{}l@{}}\textbf{Computer}\vspace{-0.4em}\\\textbf{Vision}\end{tabular}}     & Hough Transform \cite{2021hosna}     & Sub-inner     &Imperfect nested     \\
    % \multirow{2}{*}{\begin{tabular}[c]{@{}l@{}}\textbf{Computer}\\\textbf{Vision}\end{tabular}}      &\multirow{2}{*}{\begin{tabular}[c]{@{}l@{}}{Hough\\Transform}\end{tabular}}
    % & \multirow{2}{*}{Sub-inner}  &\multirow{2}{*}{\begin{tabular}[c]{@{}l@{}}{Imperfect \\Vision}\end{tabular}}
    \hline
    
    \multirow{7}{*}{\begin{tabular}[c]{@{}l@{}}\textbf{Mobile}\vspace{-0.4em}\\\textbf{Communication}\end{tabular}}     & CRC \cite{2001mibench}    & Innermost     & {\begin{tabular}[c]{@{}c@{}}Imperfect nested\vspace{-0.4em}\\Serial Loops \end{tabular}}      \\
                                                                                                         \cline{2-4}
                                                                                                         & \multirow{2}{*}{ADPCM \cite{2001mibench} }  &\multirow{2}{*}{ Serial branches }    & \multirow{2}{*}{N/A }    \\ \\
                                                                                                         \cline{2-4}
                                                                                                         & SC Decode \cite{arikan2009channel}  & Innermost     & {\begin{tabular}[c]{@{}c@{}}Imperfect nested\vspace{-0.4em}\\Serial Loops \end{tabular}}    \\
                                                                                                         \cline{2-4}
                                                                                                         & LDPC Decode \cite{richardson2008modern}  & {\begin{tabular}[c]{@{}c@{}}Nested branches\vspace{-0.4em}\\Innermost\end{tabular}}    & {\begin{tabular}[c]{@{}c@{}}Imperfect nested\vspace{-0.4em}\\Serial Loops \end{tabular}}      \\
    \bottomrule[2pt]
    \end{tabular}
    }
    \vspace{-22pt}
\end{table}

%% file: MICRO/2_Background.tex
\section{Background}\label{sec:2}

\begin{figure}[t]
\setlength{\abovecaptionskip}{3pt}
  \centering
  \includegraphics[width=1\linewidth]{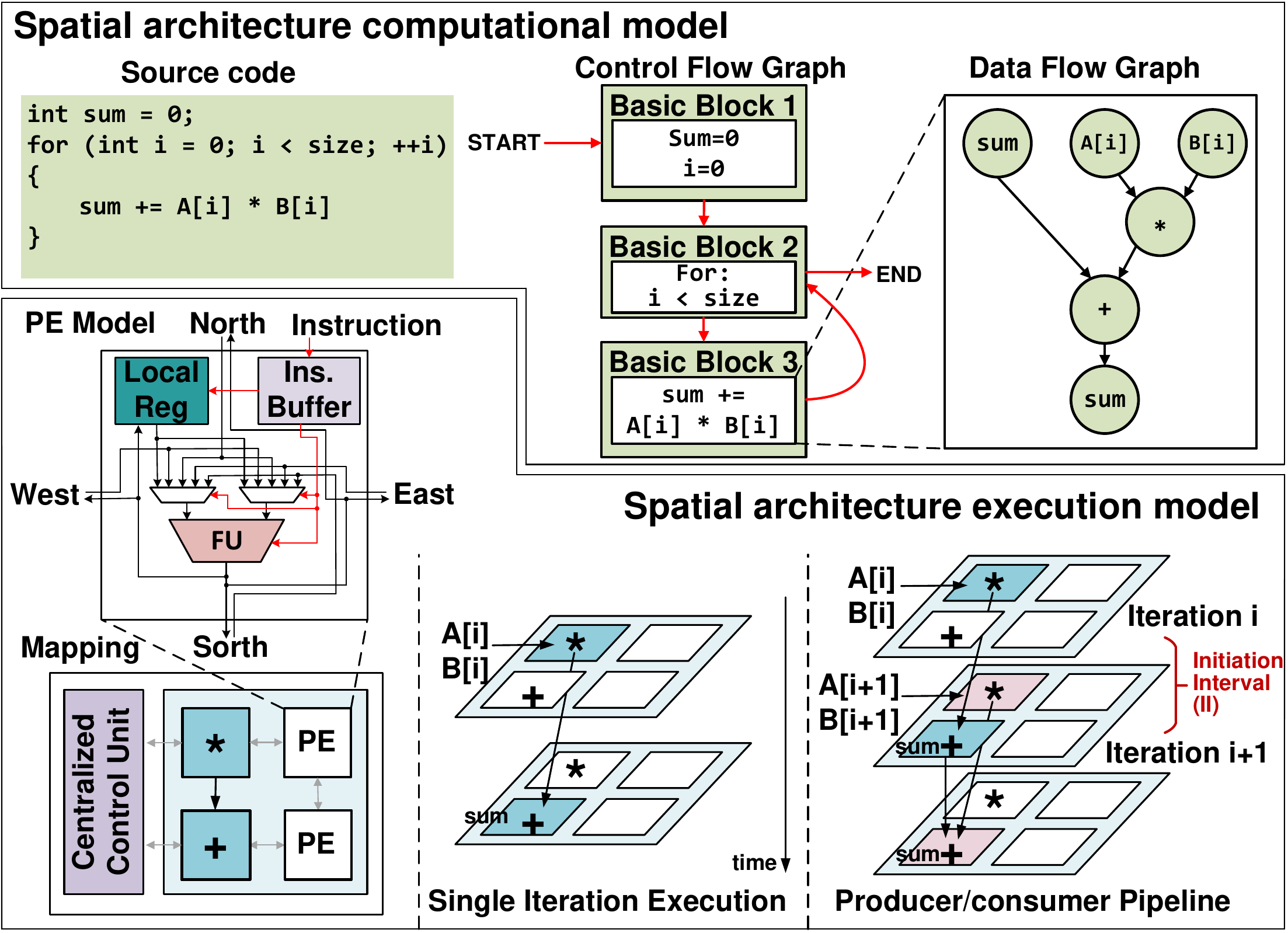}
  \caption{Spatial Architecture Computational Model and Execution Model.}
  \label{fig:fig2_1}
  \vspace{-15pt}
\end{figure}

\begin{figure*}[t]
%\vspace{-18pt}
\setlength{\abovecaptionskip}{0pt}
  \centering
  \includegraphics[width=1\linewidth]{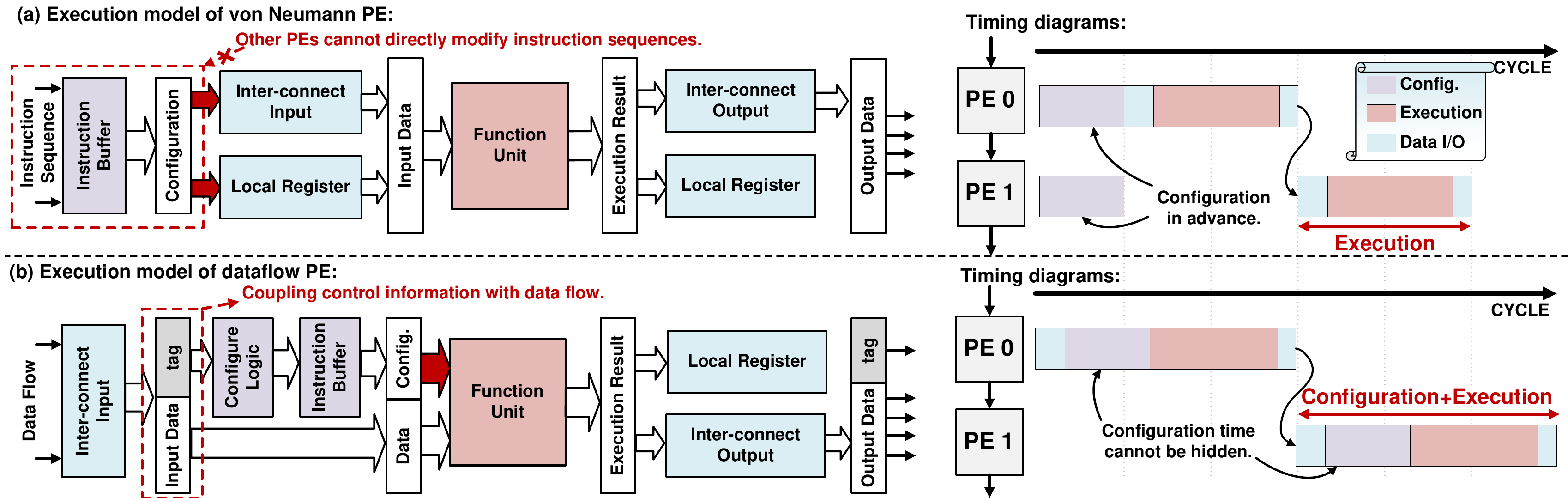}
  \caption{PE execution model categories: von Neumann PE and dataflow PE.}
  \label{fig:fig2_2}
  \vspace{-15pt}
\end{figure*}

%开文介绍
This paper presents a challenge with the intensive control flow algorithm employed in Spatial Architectures (SAs). To provide a comprehensive understanding of this problem, we first provide an overview of the SA's computational model and execution model.

\input{MICRO/CGRAtax}

\subsection{SA Computational Model}\label{subsec:2.1}

%介绍SA的计算模型是CDFG，其中CFG和DFG以及BB的概念都是什么
Upper part of Figure \ref{fig:fig2_1} shows the computational model of SA. A SA normally uses a Control Data Flow Graph (CDFG) as its computational model, where a program running on the SA is represented as CDFGs. The CDFG consists of a control flow graph (CFG) and data flow graphs (DFGs). A DFG \cite{1974DFG} is a graph that depicts operations as nodes and data dependencies as edges. As there is no control dependencies in DFG, it is usually embedded in a basic block (BB). The BB has a single entry and a single exit. The CFG \cite{1970CFG} is a graph whose nodes are BBs, and the edges represent the control dependencies between the BBs.

\subsection{SA Execution Model}\label{subsec:2.2}
%SA的并行性执行方式，生产消费并行，映射的过程。该模型是符合阿姆达尔定律的
%介绍SA的架构特征，基本工作原理
We briefly introduce the essential architecture of SA and then elaborate the hardware execution model at array-level and PE-level. We believe that understanding the root cause of control flow handling problem is worth another re-examination of the hardware execution model.

\noindent\textit{\textbf{SA Hardware Architecture:}} 
As shown in the lower left of Figure \ref{fig:fig2_1}, a SA consists of a group of processing elements (PEs) interconnected by an on-chip network. A PE typically includes a set of functional units, such as adders, multipliers, and shifters. These PEs are designed to support higher-level operations such as multiplication. Moreover, the PEs can be reconfigured to perform different tasks, and the interconnect network can be programmed to support various data flows and communication patterns, allowing the PEs to be connected in different ways to form various computational structures.

%first shows the essential hardware architecture of SA, which consists of processing elements (PEs) and the  It typically includes a set of functional units, such as adders, multipliers, and shifters, that are connected by a configurable interconnect network. The interconnect network can be programmed to support various data flows and communication patterns, allowing the PEs to be connected in different ways to form various computational structures. 
\noindent\textit{\textbf{SA Array-level Execution Model:}} 
SA needs programming to run applications. A typical way is to map the DFGs of the program onto the PE array, along with a set of hardware resources that will be used to execute the tasks. The hardware resources can include functional units, interconnects, and memory blocks. A mapping algorithm is then used to determine the optimal placement of tasks on the PEs and to configure the interconnect network to support the required data flows. Such configurations of PE and networks are achieved through instructions. 

The producer/consumer pipeline model is a crucial characteristic of SA array-level execution model that enables efficient data transfer and computation. Each PE is assigned a specific operator in the DFG, and multiple PEs are spatially interconnected to form a pipeline. Figure \ref{fig:fig2_1} lower right part depicts the producer/consumer pipeline, wherein the pipeline initiation interval (II) equals 1, which means that for each cycle, a new loop iteration can begin. As a result, the producer/consumer pipeline model affords two crucial benefits: high parallelism and effective hardware resource utilization.

%\textcolor{red}{Note that the control flow plays key role in this model. We will delve into PE-level execution model to pinpoint XXX factors. TODO}

\begin{figure*}[!t]
%\vspace{-18pt}
\setlength{\abovecaptionskip}{0pt}
  \centering
  \includegraphics[width=1\linewidth]{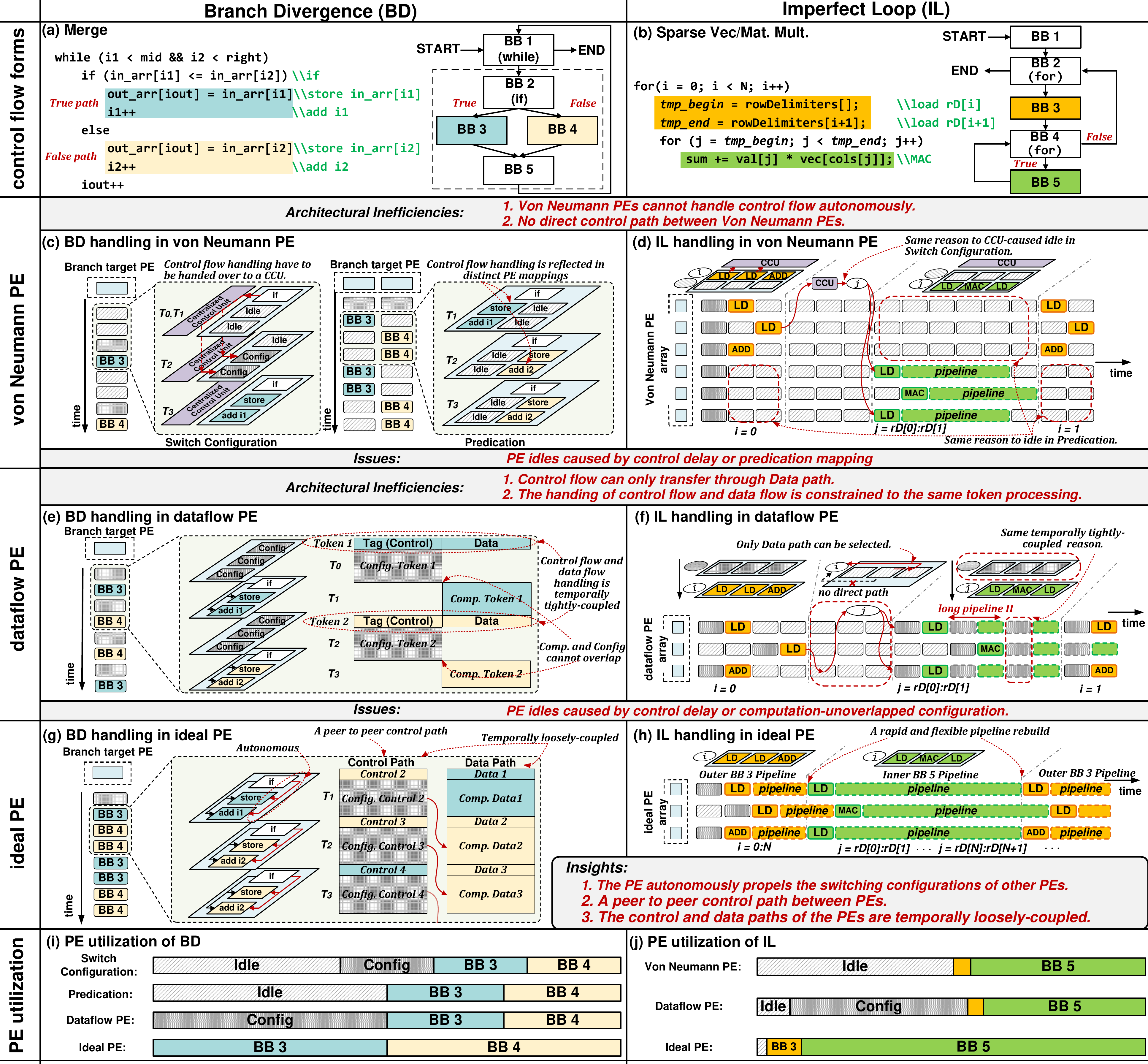}
  \caption{Von Neumann PEs and dataflow PEs face control flow handling challenges when dealing with two typical control flow forms: Branch Divergence and Imperfect Loop.}
  \label{fig:fig3_1}
  \vspace{-15pt}
\end{figure*}

\subsection{PE Execution Model}\label{subsec:2.3}
%进一步地，SA PE的分类，冯诺依曼PE和数据流PE
In this work, our goal is to comprehensively examine the existing SA execution model and pinpoint the root cause of inefficient control flow handling both at the array-level and PE-level. To achieve this goal, we first conduct a comprehensive survey of SAs in the past decade, as shown in Table \ref{table:tab8}. We then categorize them into von Neumann architecture-derived \cite{von1993first} and dataflow architecture-derived \cite{dataflow1983tagged} PE according to their control flow handling schemes, as shown in Figure \ref{fig:fig2_2}\footnote{The related work is detailed in Section \ref{sec:9}}. We assume that configuring a PE takes one cycle, and executing an instruction takes two cycles (This does not necessarily indicate an accurate timing for all architectures, but only show a relative time cost all over this paper). We will delve into the distinct ways they employ to handle control flow and identify their respective limitations.

%observe that PE execution models are derived from two classic architecture, namely, the von Neumann architecture \cite{von1993first} and the dataflow architecture \cite{dataflow1983tagged}. 
%We categorize them into von Neumann PE and dataflow PE, as shown in Figure \ref{fig:fig2_2}\footnote{The related work is detailed in Section \ref{sec:9}}. 
%\textcolor{red}{unfortunately, we find both execution models falls short of handling control flow transfer.}

%冯PE是在执行前先从指令buffer里面取到配置来配置其他模块。传统的冯诺依曼架构的控制流实际是指令的执行顺序，其通过PC指针表达。在演变而来的冯诺依曼PE中，PC指针的角色一般由有限状态机或控制核代替。切换配置的逻辑也是会按照编译时预先设定的进程快速重新配置。冯诺依曼PE在执行时的各个PE的配置逻辑还是分布式的，孤立的，无法直接被其他PE改变。并且为了适配空间架构的流水线（生产者消费者并行性），每个指令可能会对一段时间的数据输入生效。
\noindent\textit{\textbf{Control Flow Handling of von Neumann PE:}} 
A typical execution model of von Neumann PE is shown in Figure \ref{fig:fig2_2} (a). Von Neumann PE gets its configurations from the instruction buffer to configure the interconnect input/output, local register, and function unit. In the traditional von Neumann architecture, the execution sequence of instructions is controlled by the program counter (PC) pointer. However, in the evolved von Neumann PE, the PC pointer is often replaced by a finite state machine or a control core. This enables more flexible control flow and quicker reconfiguration of the processor. The logic of switching configurations is pre-set at compile time and can be quickly reconfigured. During execution, each von Neumann PE has distributed and isolated configuration logic, which cannot be directly changed by other PEs. To adapt to the pipeline of the spatial architecture, each instruction may take effect on the data input for a period of time, enabling producer-consumer parallelism.

%这个是原先写的，文字我直接改进去了, which is statically decided in compiling time. The control flow is represented via instruction sequences for von Neumann PE. Then the operation sequence is executed with the data stream being fed in. Unlike the traditional von Neumann architecture that includes a PC controller, von Neumann PE does not have PEA-wise real-time instruction control. The control information (instruction sequences) is distributed and statically determined during the compilation phase, as shown in the timing diagram. Therefore, von Neumann PE cannot receive control information from other PEs at runtime. Modifying the configuration needs to be done via the core with the configuration bus or the control network.

%数据流PE是根据数据选择buffer中的指令配置，然后执行。因此有两个特点：1.数据流PE在执行时，PE的配置逻辑不是孤立的，可以接收其他PE的调控，他们实现这种调控的方式是保持数据和tag捆绑成一个token，为了保证token一致性，每一次执行都是数据等tag，新tag等新数据。2.由于这种一致性，一条指令仅会对与之匹配的数据生效，而非一组流水线。
%

\noindent\textit{\textbf{Control Flow Handling of dataflow PE:}} Figure \ref{fig:fig2_2} (b) displays a standard model for executing dataflow PEs. A token is used as input, consisting of a data and a tag. The tag activates the configuration, while the data performs the operation. This means that the dataflow PE's configuration can be altered by other PEs while in use. However, the token links control flow and data flow, leading to some limitations. Each PE execution requires configuration before execution, causing latency overhead. Moreover, this coupling limits the effect of an instruction to the current data, hampering pipeline control.

%% file: MICRO/CGRAtax.tex
\begin{table}[!b]\Huge
   \vspace{-15pt}
    \setlength{\abovecaptionskip}{0pt}
        \renewcommand{\arraystretch}{1.1}
        \centering
        \caption{SA taxonomy by PE execution model.}
        \label{table:tab8}
        \resizebox{1\linewidth}{!}{%
        \begin{tabular}{lll} 
        \toprule[3pt]
                                                                  &{\centering\hspace{-40pt}} \textbf{Architecture} & \textbf{Mechanism for configuration triggering }                               \\ 
        \hline
        \multirow{10}{0.148\linewidth}{\hspace{0pt}\rotatebox{90}{\textbf{von Neumann PE}}} & RICA \cite{2007RICA}         & A core processor that generates the overall configuration signal.     \\
                                                                  & DRP \cite{DRP}          & Switching all PE configurations via a finite state machine.           \\
                                                                  & DySER \cite{2012dyser}        & Configuration update via external processor signal.                   \\
                                                                  & FPCA \cite{2014FPCA}         & External processor assignments.                                       \\
                                                                  & DORA \cite{2016DORA}         & A counter determines the end and update of the configurations.        \\
                                                                  & Plasticine \cite{2017plasticine}   & A counter controls the distribution and execution of configurations.  \\
                                                                  & Softbrain \cite{2017stream}   & Processor fetches instruction from memory.                       \\
                                                                  & SPU \cite{2019spu}         & Processor fetches instruction from memory.                       \\
                                                    
                                                                  & MP-CGRA \cite{MP-CGRA}     & Distributed instruction counters.                                    \\ 
                                                                  & DRIPS \cite{2022drips}       & The centralized controller dynamically changes the map table.         \\
                                                                  & RipTide \cite{gobieski2022riptide}       & Processor fetches instruction.          \\
        \hline
        \multirow{7}{0.148\linewidth}{\hspace{0pt}\rotatebox{90}{\hspace{12pt}\textbf{dataflow PE}}}  & TRIPS \cite{2004trips}        & An instruction window to determine instruction execution.             \\
                                                                  & Wavescalar \cite{2003wavescalar}  & According to the data, configurations are fetched to execute.         \\
                                                                  & TIA \cite{TIA}         & Scheduler selects intructions based on the input data.                \\
                                                                  & T3 \cite{2013T3}          & An instruction window to determine instruction execution.             \\
                                                                  & SGMF \cite{2014sSGMF}        & The corresponding thread is executed when the token arrives.           \\
                                                                  & dMT-CGRA \cite{2018dMT-CGRA}    & An instruction window to determine instruction execution.             \\
                                                                  
        \bottomrule[3pt]
        \end{tabular}
        }
    \end{table}

%% file: MICRO/3_Motivation.tex
\section{Challenges and Motivations}\label{sec:3}

%Based on the basic introduction of control flow handling methods of von Neumann PE and dataflow PE, we detailedly describe how these two PEs execute control flow-intensive kernels using two representative control flows, namely \textit{branch divergence} and \textit{imperfect loop}, which are adequate for modeling a wide range of algorithms. The average percentage of the two forms can be over 40\% \cite{yin2021subgraph} in MachSuite \cite{2014machsuite}, Rodinia \cite{che2009rodinia} and PolyBench \cite{polybench}. 

In this section, we will provide a detailed explanation of how von Neumann PE and dataflow PE handle control flow when executing control flow-intensive kernels. We will use two representative control flows, namely \textit{Branch Divergence} and \textit{Imperfect Loop}, which are commonly found in various algorithms. These control flows account for over 40\% of the average percentage in popular benchmark suites such as MachSuite \cite{2014machsuite}, Rodinia \cite{che2009rodinia} and PolyBench \cite{polybench}.
    
Our goal is to show the root cause of existing spatial architectures' awkwardness in handling the control flow, which is due to the lacking mechanism for \textbf{\textit{autonomous}}, \textbf{\textit{peer-to-peer}}, and \textbf{\textit{temporally loosely-coupled}} control flow information transfer among PEs in SA. This observation motivates us to propose a decoupled control flow plane for SA, which includes redefined PE architecture and features.

\subsection{Two Typical Control Flow Forms}\label{subsec:3.1}

\textbf{\textit{Branch Divergence}} is a prevalent issue in CFGs, where the program's control flow divides into different execution paths due to the presence of conditional branches. This occurs when the program encounters a decision point, where it must choose between multiple execution paths based on specific conditions. Branch Divergence is common in various kernels such as \textit{Sort} and \textit{Merge}, in the database and sparse computing. As an example, Figure \ref{fig:fig3_1} (a) shows a code snippet in Merge, where the data flow within the conditional branch dynamically forks into two paths (true and false), or BBs. This results in divergent execution paths and can lead to poor PE utilization in both existing PE execution models. 
        
\textbf{\textit{Imperfect Loop}} is another typical control flow form which can be characterized as nested loops, with computations present in the outer loop bodies. It is a common feature in scientific computing and finite element analysis, particularly in applications such as blocked matrix multiplication (GEMM) and computational fluid dynamics (CFD). Figure \ref{fig:fig3_1} (b) shows a code snippet of Sparse Vec/Mat. Mult. (SPMV), where the inner loop (in green) executes every block size times while the outer loop (in yellow) executes only once. Different nested loops have varying BB execution frequencies, which can cause an imbalanced pipeline and poor PE utilization in existing PE execution models.

\subsection{Control Flow Handling Inefficiency in von Neumann PE}\label{subsec:3.2}
\input{MICRO/SOTA} 

The von Neumann PE's execution model exhibits two limitations in handling control flow: firstly, \textbf{\textit{it is unable to autonomously initiate configuration changes in other PEs}}; secondly, \textbf{\textit{there exists no direct channel for the transmission of control information between PEs}}. Consequently, in the face of Branch Divergence, von Neumann PEs typically resort to two implementation approaches.

\textit{\textbf{Branch Divergence: Switch Configuration}} 
The first method (shown in the left half of Figure \ref{fig:fig3_1}  (c)) is to implement the branch in the fashion of switching PE configurations in the time dimension. Specifically, the branch target PEs need to wait for the control flow information and switch the configurations. Due to the inability of the branch PE to autonomously modify the control information of the target PEs, and the lack of a direct channel for control flow information transmission between them, the branch PE is constrained to indirectly convey control information through an external Centralized Control Unit (CCU). This approach is clumsy. As the branch PE needs to transfer the control flow result to the CCU. Then the CCU replies to the branch target PE through the configuration network \textit{while the whole array is left idle}.

\textit{\textbf{Branch Divergence: Predication}}
The second approach, referred to as "Predication" (shown in the right half of Figure \ref{fig:fig3_1} (c)), is more prevalent. It involves converting branches into distinct data paths in spatial dimension by consuming additional PEs. The configurations for both branch targets are pre-configured in two target PE lanes, respectively. Subsequently,  the correct PE lane is selected from these two paths for the following BB according to the branch result. However, the not taken PEs will be left idle. \textit{It would be great if this idle resource could be used as other kernels}.

\textit{\textbf{Imperfect Loop:}} 
The CFG of Imperfect Loop can be seen as a variation of Branch Divergence. BB 4 is responsible for making a branch decision, with one branch leading to BB 5 and the other to BB 2 and BB 3. As a result, the von Neumann PE typically employs the Predication method in Branch Divergence. Based on this, there is a key point in the SPMV algorithm: the outcome of BB 3 determines the loop boundary for BB 5. Thus, it bears resemblance to the Switch Configuration employed in Branch Divergence. In this case, BB 3 transmits the control flow to the Centralized Control Unit, which subsequently configures the loop generator of the inner loop. As shown in Figure \ref{fig:fig3_1} (d), \textit{the low utilization rate of Von Neumann PE in various stages is quite significant, and it can be attributed to a reason similar to the idle state under Branch Divergence.}
        
\begin{figure*}[t]
    %\vspace{-18pt}
    \setlength{\abovecaptionskip}{0pt}
    \centering
    \includegraphics[width=1\linewidth]{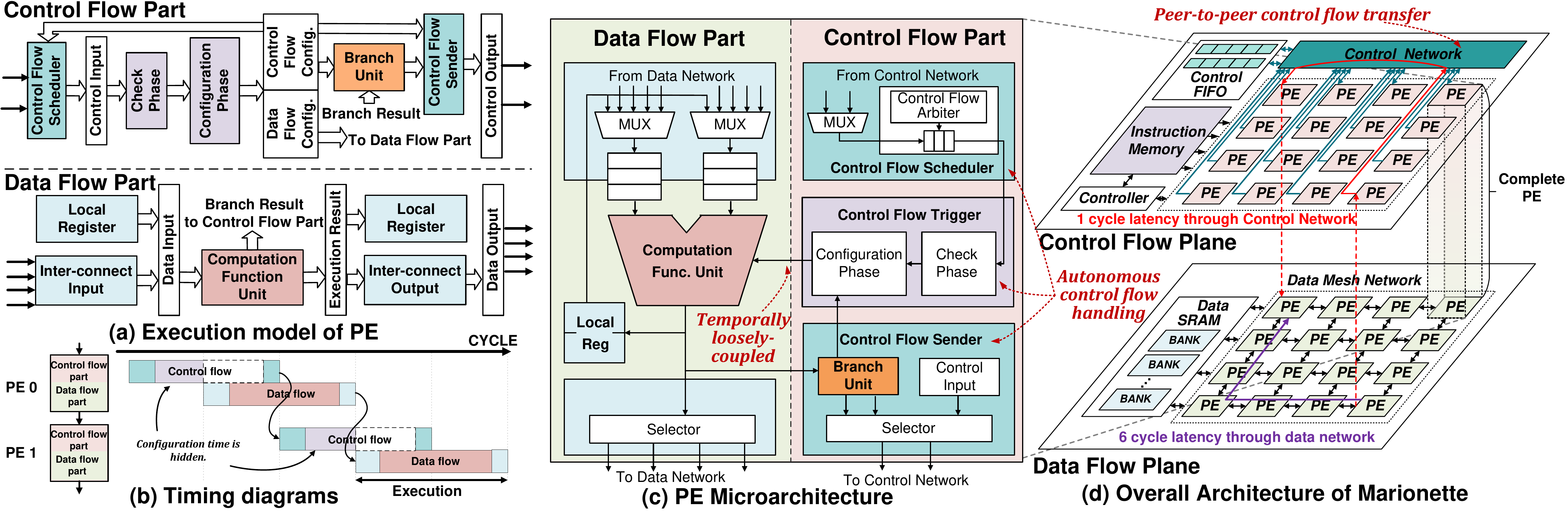}
    \caption{Marionette PE execution model, timing diagrams, PE micro-architecture, and overall architecture.}
    \label{fig:fig4_1}
    \vspace{-15pt}
\end{figure*}          

\subsection{Control Flow Handling Inefficiency in Data flow PE} \label{subsec:3.3}

The execution model of the dataflow PE demonstrates a restriction in managing control flow. \textbf{\textit{The constraint stems from the binding of control and data flow within tokens. This close-knit coupling in both temporal and spatial dimensions restricts the control flow transferring.}}

\textit{\textbf{Branch Divergence:}}
We expose the challenge caused by the temporal tight coupling between control flow and data flow when dataflow PEs perform Branch Divergence, as shown in Figure \ref{fig:fig3_1} (e). The concurrent arrival of both tag (control flow information) and data flow via the same channel at the same time necessitates a PE configuration stage as a consequent operation of data entry, leading to an explicit overhead for PE configuration. \textit{Unfortunately, this explicit overhead results in a suboptimal utilization of the PE.}

%However,  the tag (control flow information) and data flow arrive via the same channel and at the same time. Therefore, a PE configuration stage has to be the consequent operation of data entry, thus making PE configuration an explicit overhead. This explicit PE configuration overhead results in low PE utilization. 

\textit{\textbf{Imperfect Loop:}}
When executing an Imperfect Loop, dataflow PEs can accentuate the challenges arising from the close coupling of control flow and data flow, manifesting in both temporal and spatial domains, as shown in Figure \ref{fig:fig3_1} (f). On the one hand, the synchronization of control flow and data flow naturally leads to a longer pipeline II, as elaborated in the preceding paragraph. On the other hand, the close spatial couping of control flow and data flow implies that control information transferring reliant on the data path is frequently inflexible. For instance, in the absence of a direct control flow pathway between the second PE and the loop generator, control flow information must traverse the red data path, leading to pipeline idleness. \textit{Consequently, even though the dataflow PE has some autonomy, the inherent drawback results in a substantially reduced utilization rate. }

\subsection{Insights and Motivations}\label{subsec:3.4}

We can observe that both von Neumann PE and dataflow PE will cause significant PE idleness, as shown in Figure \ref{fig:fig3_1} (i)(j). This is mainly due to (1) PEs cannot autonomously change the configuration of other PEs;  (2) Control flow transmission between PEs is restricted, manifested in both spatial and temporal dimensions: 1. Spatially, control flow transmission is not direct, but instead realized by switching configurations through centralized control units or coupled into the data flow (i.e. tag or predication). 2. Temporally, control flow transfer is also constrained. The configuration stage is hard to overlap with computation since the configuration function and computation function are temporally tightly-coupled.

Facing the intensive control flows, an ideal PE is expected to (1) be able to \textit{\textbf{autonomously change the control information of other PEs}}, (2) incorporate a \textit{\textbf{peer-to-peer control flow path}} to enable agile control information transmission, and (3) feature \textit{\textbf{an temporally loosely-coupled control flow path}} to facilitate the concurrent execution of current-stage computation and next-stage configuration within the same stage, as shown in Fig. \ref{fig:fig3_1} (g)(h). To achieve this objective, a critical requirement is to \textit{\textbf{decouple the control flow handling and dataflow handling}}, which calls for a new architectural scheme. Our architecture is predicated on this assumption, and the superior performance of our design relative to other state-of-the-art architectures is demonstrated in Table \ref{table:SOTA}.

%% file: MICRO/SOTA.tex
\begin{table}[t]\Large
\setlength{\abovecaptionskip}{0pt}
\renewcommand{\arraystretch}{1.2}
\centering
\caption{Comparison of Marionette's ability to handle control flow with other state-of-the-art architectures}
\label{table:SOTA}
\vspace{3pt}
\resizebox{1\linewidth}{!}{%
\begin{tabular}{ccccccc} 
\toprule
                                                                                                    & \textbf{Softbrain} & \textbf{TIA} & \textbf{Dyser} & \textbf{Plasiticine} & \textbf{RipTide} & \textbf{Marionette (ours)}  \\ 
\hline
\begin{tabular}[c]{@{}c@{}}\textbf{Autonomously}\\\textbf{control other PEs}\end{tabular}         & \LARGE\ding{53}                 & \LARGE\ding{51}          & \LARGE\ding{53}             & \LARGE\ding{53}                   & \LARGE\ding{53}               & \LARGE\ding{51}                         \\ 
\hline
\begin{tabular}[c]{@{}c@{}}\textbf{Peer-to-peer}\\\textbf{control flow path}\end{tabular} & \LARGE\ding{53}                 & \LARGE\ding{53}           & \LARGE\ding{53}             & \LARGE\ding{53}                   & \LARGE\ding{53}               & \LARGE\ding{51}                         \\ 
\hline
\vspace{-2pt}
\begin{tabular}[c]{@{}c@{}}\textbf{Temporally loosely-}\\\textbf{coupled with dataflow}\end{tabular}           & \LARGE\ding{53}                 & \LARGE\ding{53}           & \LARGE\ding{53}             & \LARGE\ding{53}                   & \LARGE\ding{53}               & \LARGE\ding{51}                          \\ 

\bottomrule
\end{tabular}
}
\vspace{-15pt}
\end{table}

%% file: MICRO/4_Innovation_Architecture.tex
\section{Design}\label{sec:5}

%第三章已经把问题解释出来了：
%SA需要一种新的机制，可以把众多PE的控制流部分单独设计优化，并且形成一个新的抽象层面，达到效果：众多PE控制流进行高效的交互。%The imperative for a novel feature within the SA architecture that can singularly design and optimize the control flow of several processing elements (PEs) cannot be overstated. To this end, the introduction of a new level of abstraction that fosters the efficient interaction of multiple PE control flows is indispensable. Therefore, this paper proposes an innovative feature that will enable the individualized design and optimization of control flow for multiple PEs, while creating a new abstraction level to facilitate their interaction.
%第三章motivate了：一是SA的架构应该重新设计。二是pe在处理控制流的时候应该有机制。三是PEA层级协调需要有一个机制。文字没写
The current PE execution models expose a deficiency in control flow handling ability. It becomes imperative to deploy an autonomous, peer-to-peer and temporally loosely-coupled control channel. We propose that in spatial architectures, control flow handling should be separated from the current hybrid designs that combine control flow handling with data flow handling. This introduces a new layer of abstraction, called the control flow plane of the spatial architecture, which facilitates the separate design and optimization of control flow handling for each PE. This approach represents the only viable means to swiftly reconfigure a set of PEs and achieve control coordination within the spatial architecture. We introduce Marionette, a spatial architecture design with decoupled control flow plan and data flow plane. Specifically, we architect a control flow plane for existing spatial architecture, which incorporates three innovative features. This section presents our proposal for Marionette and organizes the discussion to describe its key features.

\begin{enumerate}
\item What is the control flow plane of Marionette, and how to realize a autonomous, peer-to-peer and temporally loosely-coupled control flow handling capability by establishing a control flow plane?
\item How is the Proactive PE Configuration used to achieve timely and computation-overlapped configuration through the Control Flow Sender, and what solutions does it offer for Branch Divergence and pipeline initiation?
\item How does the Agile PE Assignment enhance the pipeline performance of Imperfect Loops through a refined Marionette scheduling strategy and Control Flow scheduler?
\end{enumerate}

%We have explored that spatial architectures call for control plane designs to ensure the consistency of the execution model and computational model (i.e., maintaining explicit control flow path and data flow path in both execution model and computational model). we summarize the design requirements for the control plane and data plane of the execution model based on the characteristics of the CFG and DFG in the computational model in Table \ref{table:tab3}.

\subsection{Control Flow Plane}\label{subsec:4.1}

\begin{figure}[t]
%\vspace{-15pt}
\setlength{\abovecaptionskip}{0pt}
  \centering
  \includegraphics[width=1\linewidth]{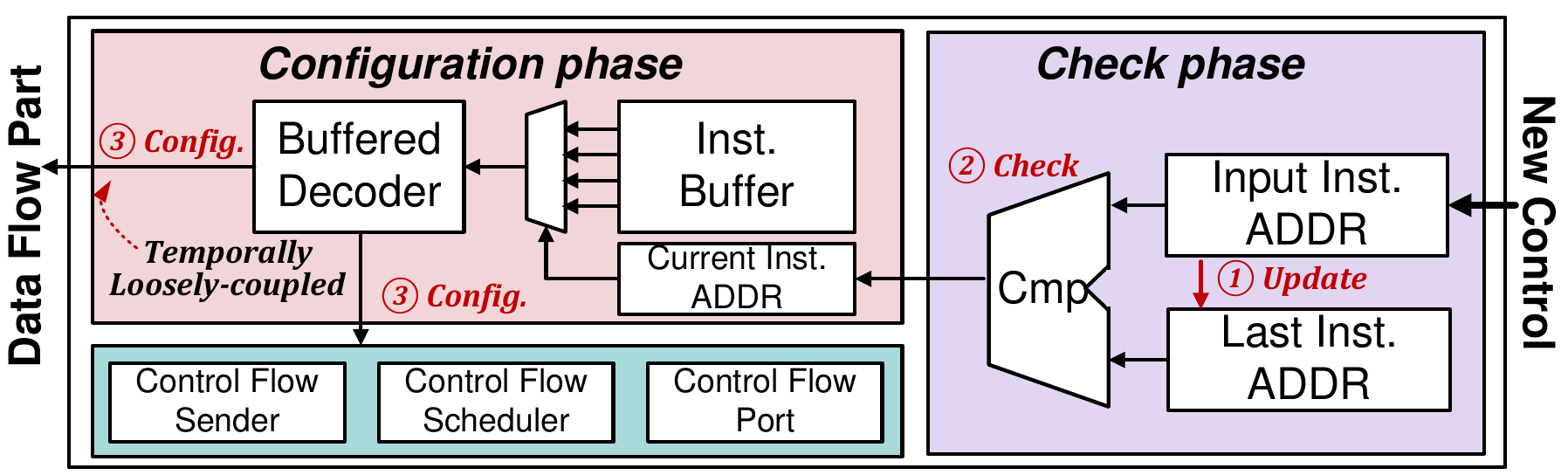}
  \caption{PE control flow trigger.}
  \label{fig:fig4_2}
  %\vspace{-15pt}
\end{figure}

%Marionette的控制面如何被抽象，都包含什么，控制网络+PE
\noindent\textbf{Control flow plane abstraction in Marionette:} As shown in Figure \ref{fig:fig4_1} (d), to attain a fully decoupled control flow plane, we have encapsulated all control flow-related components, including the Controller, Control Network, Control FIFO, and the control flow part within PEs, within Marionette's control flow plane. The primary objective of the control flow plane is to establish a correlation between CFG and the hardware implementation. \textbf{\textit{In this process, the control flow is represented by instruction addresses, and the PE generates and sends new instruction address to other PEs to realize an autonomous control flow handling. }}A cluster of PEs operating on a consistent instruction address can depict a BB. Similarly, Marionette's data flow plane encompasses components such as the Data network, Data SRAM, and Data flow part inside the PE. Upon receiving the corresponding configuration from the control flow part, the data flow plane is responsible for performing data flow computations and accessing memory, thus ensuring the realization of the DFG.

\noindent\textbf{Execution model of Marionette PE:} In order to achieve autonomous, peer-to-peer, and temporally loosely-coupled control flow handling within PEs, we decouple and optimize the control flow component of the conventional PE execution model. The Marionette PE execution model, illustrated in Figure \ref{fig:fig4_1} (a), demonstrates the decoupling of the control flow part and data flow part. We design the micro-architecture of three control flow  components: the Control Flow Scheduler, Control Flow Trigger, and Control Flow Sender, along with a corresponding ISA that enables independent control flow handling and ports. This permits the free transmission and receipt of control flow, unrestricted by the data flow within the PE.

%Control Flow Trigger放着了，不知道行不行，我想了两种角度去说，一个是介绍它有一个避免频繁切换配置的能力，但是和性能搭不上边。还有就是可以介绍配置组成成分时候引到这个，没太想好怎么说

Control Flow Trigger, shown in Figure \ref{fig:fig4_2}, is the pivotal configuration unit of the Marionette PE framework. It is composed of two phases, namely the check phase and the configuration phase. The Control Flow Trigger is designed to sustain the configuration determined in the configuration phase until a fresh control input is detected during the check phase. This contrasts with data flow PE instructions, which are solely responsible for a single calculation. By virtue of the majority of PEs executing within the confines of the same basic block's producer-consumer pipeline, the Control Flow Trigger obviates the overhead of switching instructions.

\noindent\textbf{Autonomous control flow handling:} As previously noted, the autonomy of the Marionette PE stems from its ability to generate instruction addresses for subsequent PEs. The configuration of control flow part provides a range of instruction addresses. 

\begin{figure}[b]
%\vspace{-18pt}
\setlength{\abovecaptionskip}{0pt}
  \centering
  \includegraphics[width=1\linewidth]{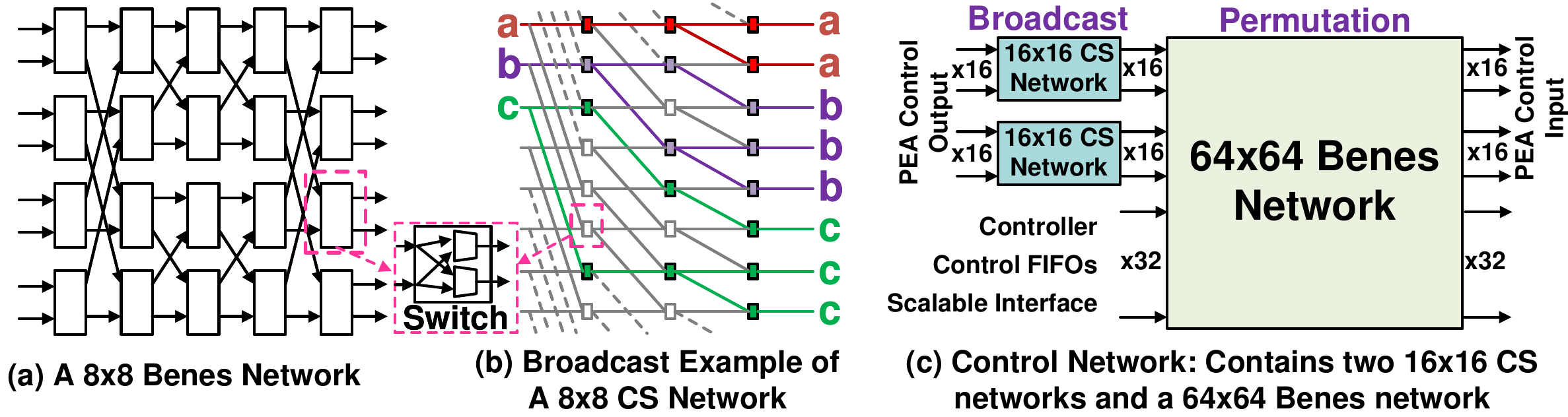}
  \caption{A dedicated control network enables peer-to-peer transfer of control flow between PEs.}
  \label{fig:fig4_4}
%\vspace{-15pt}
\end{figure}

\noindent\textbf{Temporally loosely-coupled control flow handling:} 
Given that the Control Flow Scheduler, Control Flow Trigger, and Control Flow Sender on the control flow plane possess distinct instruction sets and execution procedures, the control flow and data flow are inherently decoupled. Consequently, it allows for overlapping of the next configuration phase with the current computation phase through Proactive PE Configuration, as detailed in Section \ref{subsec:4.2}. This approach facilitates the transfer of control and data flow between PEs, as demonstrated in Figure \ref{fig:fig4_1} (b).

\noindent\textbf{A peer-to-peer Control Network:} To enable timely and flexible peer-to-peer control flow transfer with minimal area overhead, we design a control network based on the Benes network. This well-known rearrangeable non-blocking network \cite{1962BENESNET} has a butterfly-shaped interconnection structure, and a much smaller number of node switches than the Crossbar network, serving as our design starting point due to its low area overhead and high flexibility (Figure \ref{fig:fig4_4} (a)). However, it lacks broadcasting capabilities. To address this, we incorporate the Consecutive Spreading (CS) network \cite{1988CSNET}, which performs broadcast and has a smaller area overhead than cascading multiple same-sized networks. We present a CS-Benes network that connects PEs, control FIFOs, and the controller, providing configurable network output with a fixed connection and no arbitration during control transfers. Each path in the network contributes one element of throughput every cycle. We reserve many extensible interfaces. Figure \ref{fig:fig4_4} (c) displays the specific interface design of our CS-Benes network, and in Section \ref{subsec:8.2}, we evaluate the control network's scalability.

\begin{figure}[t]
%\vspace{-18pt}
\setlength{\abovecaptionskip}{0pt}
  \centering
  \includegraphics[width=1\linewidth]{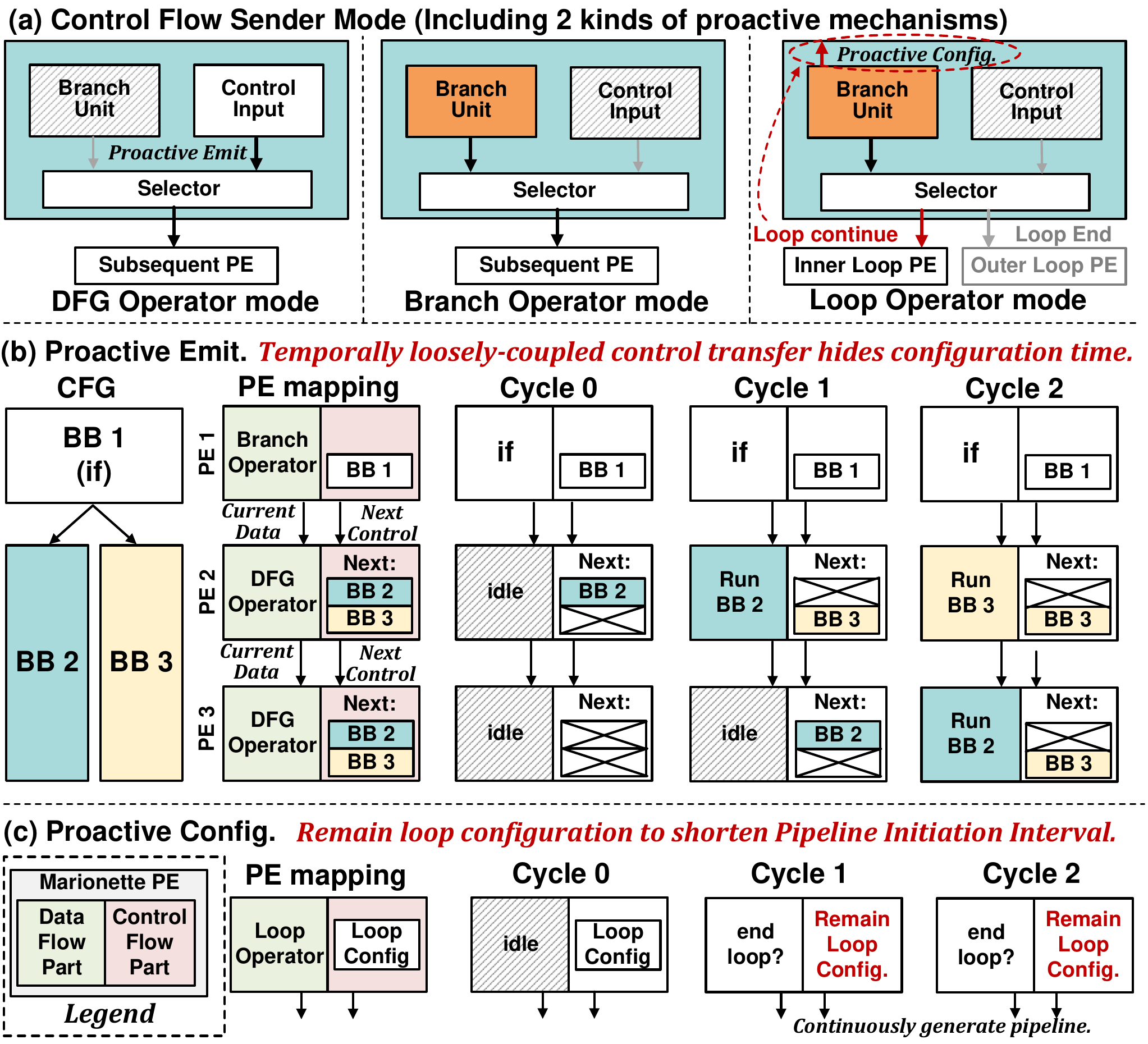}
  \caption{Control Flow Sender and Proactive PE Configuration feature for Branch Divergence and Pipeline Initiation.}
  \label{fig:fig4_3}
\vspace{-15pt}
\end{figure}

\subsection{Proactive PE Configuration} \label{subsec:4.2}

\begin{figure*}[!t]
%\vspace{-18pt}
\setlength{\abovecaptionskip}{0pt}
  \centering
  \includegraphics[width=1\linewidth]{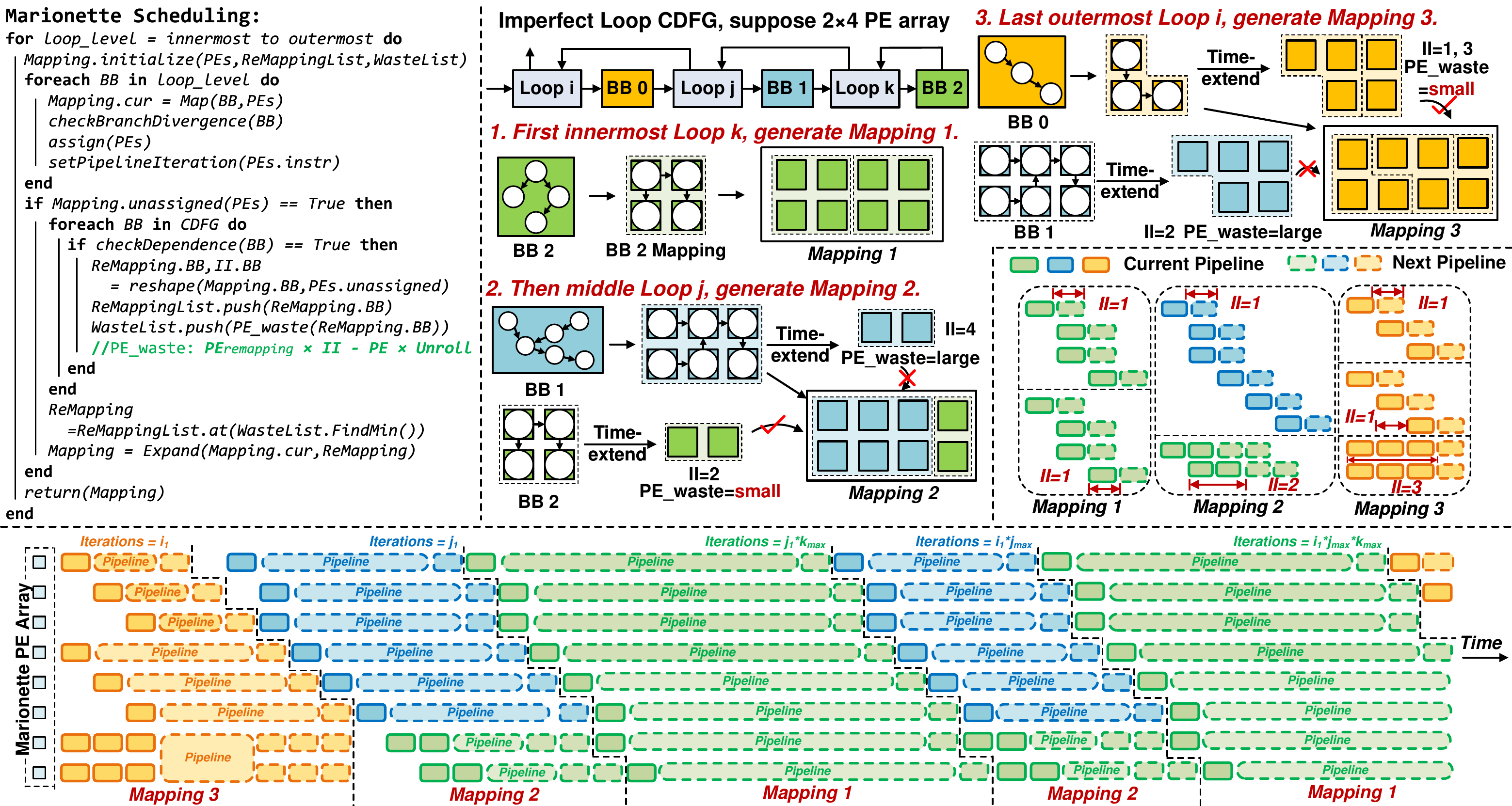}
  \caption{Agile PE Assignment, including Marionette scheduling algorithm and an example of processing Imperfect Loops.}
  \label{fig:fig4_5}
\vspace{-15pt}
\end{figure*}

%主动控制流发射器有三种执行模式，其中两种模式下有主动能力

In the Marionette control flow plane design, the control flow and data flow are loosely-coupled in the time domain, which allows for the execution of current-stage computation and next-stage configuration within the same stage. To accelerate the configuration process of subsequent PEs, we innovate Proactive PE Configuration and develop Control Flow Sender that send control information at the earliest possible time. The data flow part of the PE currently implements three distinct operating modes for the control flow transmitter: DFG operator mode, branch operator mode, and loop operator mode, as shown in Figure \ref{fig:fig4_3} (a). Additionally, the Control Flow Sender features Proactive PE Configuration in the DFG operator mode and loop operator mode that hastens the transmission of control flow.

%The Marionette PE's data flow part carries out the computation of non-branch operators in the DFG operator mode. In this mode, it signifies that the current and subsequent PEs are located in the same BB and share the same control flow. To streamline the transmission of control flow, the configuration of the current PE is proactively sent to the subsequent PE once the configuration process is completed. As a result, by the time the current PE dispatches the data flow's result to the subsequent PE, the configuration of the latter has already been finalized.On the other hand, in the branch operator mode, the PE executes the branch (not loop) operator, indicating that the current and subsequent PEs are in different basic blocks separated by a control jump. Consequently, the control flow of the PE must await the computation result of the data flow in order to determine the control information of the subsequent PE, and no proactive control flow is transmitted.Lastly, in the loop operator mode, the data flow segment executes the loop operator, and to ensure a continuous pipeline generation, the loop configuration is maintained in advance.

The Marionette PE's data flow part executes non-branch calculation operators in the DFG operator mode. This mode indicates that the current and subsequent PEs are in the same BB and share the same control flow. To expedite control flow transmission, the current PE's configuration is sent to the subsequent PE in advance once the configuration is completed. Consequently, when the current PE sends the data flow's dataflow result to the subsequent PE, the configuration of the latter has already been completed. In contrast, the branch operator mode executes the branch (not loop) operator, indicating that the current and subsequent PEs are in different basic blocks with a control jump in between. Therefore, the control flow of the PE must wait for the data flow calculation result to determine the control information of the subsequent PE, and no proactive control flow is transmitted. Lastly, in the loop operator mode, the data flow part executes the loop operator, and to ensure continuous pipeline generation, the loop configuration is maintained in advance.

In Figure \ref{fig:fig4_1} (b), we have explained the workings of the control flow plane and data flow plane within a single PE. Furthermore, we demonstrates the execution of three Marionette PEs under Branch Divergence in Figure \ref{fig:fig4_3} (b). The scenario includes BB 1, which consists of a single branch operation, and BB 2 and BB 3, which are two branch paths with a length of two. To handle this, we assign the branch operation to PE 1 and set it to Branch Operator mode. Additionally, the two branch paths are merged and assigned to PE 2 and PE 3, which are in DFG Operator mode. In each cycle, PE 1 executes the branch instruction to select the correct branch path. It then transmits the configuration signals of the branch path to the control flow plane of the downstream PE. The data flow plane of the downstream PE executes the correct branch path, meanwhile the control flow plane accepts the upstream configuration signals in advance to configure for the next cycle. This effectively hides the configuration time overhead. The Proactive PE Configuration feature allows Marionette PE to achieve a branch target PE utilization rate that is similar to that of an ideal PE.

Figure \ref{fig:fig4_3} (c), it illustrates the PE's functionality as a loop generator. When the PE is configured in Loop Operator mode, the control flow plane continuously maintains the current loop operation. It does not accept new configuration signals until the data flow plane determines that it should exit the loop. This capability allows for the continuous generation of the loop pipeline. (This illustrates the method for minimizing Pipeline II. In reality, Pipeline II is configurable.)

\begin{figure}[!b]
\vspace{-15pt}
\setlength{\abovecaptionskip}{0pt}
  \centering
  \includegraphics[width=1\linewidth]{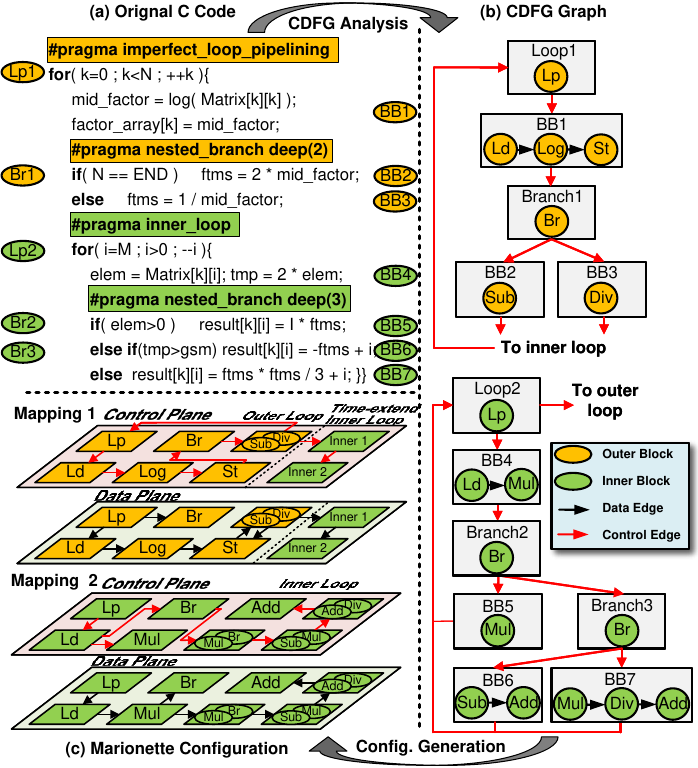}
  \caption{Example Marionette Program Transformation.}
  \label{fig:fig4_6}
%\vspace{-5pt}
\end{figure}

\subsection{Agile PE Assignment} \label{subsec:4.3}

%Based on the Marionette control flow plane design, groups of PEs are able to autonomously and quickly modify their configurations. This forms the foundation for rebalancing an imperfect loop pipeline, and we have developed Agile PE Assignment feature to improve the utilization rate of PEs in imperfect loops. Upon investigation, we identified three potential improvements that could be made to von Neumann PE and dataflow PE to enhance the low utilization rate of the imperfect loop: 1. To minimize control flow jumps following the completion of inner loop iterations, the BB of the outer loop can also establish a pipeline. 2. To ensure the execution of the inner loop consumes as many PE resources as possible. 3. To reduce the pipeline initiation interval. To this end, we have developed a loop operator (as previously discussed), and two PE assignment approaches:  1) reconfiguring the PE that originally executed the outer loop as the inner BB vectorized mapping when executing the inner loop, and 2) constructing the pipeline for the outer loop to generate control flow results in advance. 

Leveraging the Marionette control flow plane architecture, the PE exhibits autonomous, peer-to-peer and temporally loosely-coupled control flow handling capabilities, aided by a dedicated loop operator that regulates the pipeline II. This serves as the foundation for realigning the flawed loop pipeline, and the Agile PE Assignment feature we proposed enhances PE utilization in Imperfect Loops. The feature encompasses a refined Marionette scheduling strategy and a Control Flow scheduler.

The Marionette scheduling algorithm is shown in the upper left of Figure \ref{fig:fig4_5}. The frequency of BB execution varies among nested loop levels. Thus, we establish the mapping at each loop level and construct the pipeline with BB granularity. Once all BBs in the current loop level have been scheduled and unassigned PEs remain, we reshape (time-extend) the mapping of both the current layer BB and the inner layer BB, as they satisfy control dependencies in the current state. Time Extended mapping is a widely adopted technique in compilation \cite{balasubramanian2021compiler}, which entails the folding of the initial spatial domain mapping into the temporal domain, thereby reducing PE resources while also increasing the II. However, reshaping may result in idle PEs due to the diverse DFG shapes of BBs. To address this issue, we select a mapping scheme that minimizes PE waste and expand the original mapping to generate the mapping result of the current loop level. The reshaping scheme, PE waste, and scheduling results of the three-layer nested loop algorithm are illustrated in the upper right of Figure \ref{fig:fig4_5}.

The lower part of Figure \ref{fig:fig4_5} showcases the Agile PE Assignment through Marionette timeline diagrams. The Marionette control flow plane allows for a highly flexible construction of the BB pipeline, encompassing adaptable PE resources, startup time, and pipeline II. This results in significantly enhanced PE utilization. To collect the control information generated by the outer BB pipeline, we design the Control Flow Scheduler and Control FIFOs. When the inner loop BB completes a round of loop iterations, it utilizes the pre-collected outer loop BB's control information to determine whether to initiate the next loop, thereby avoiding frequent configuration switch to the outer loop BB. Furthermore, the Control Flow Arbiter inside the Control Flow Scheduler possesses the capability to arbitrate between the current execution configuration and input control, enabling dynamic adjustment of the execution priority among BBs with varying levels of nested loops.

\begin{figure}[t]
%\vspace{-15pt}
\setlength{\abovecaptionskip}{0pt}
  \centering
  \includegraphics[width=1\linewidth]{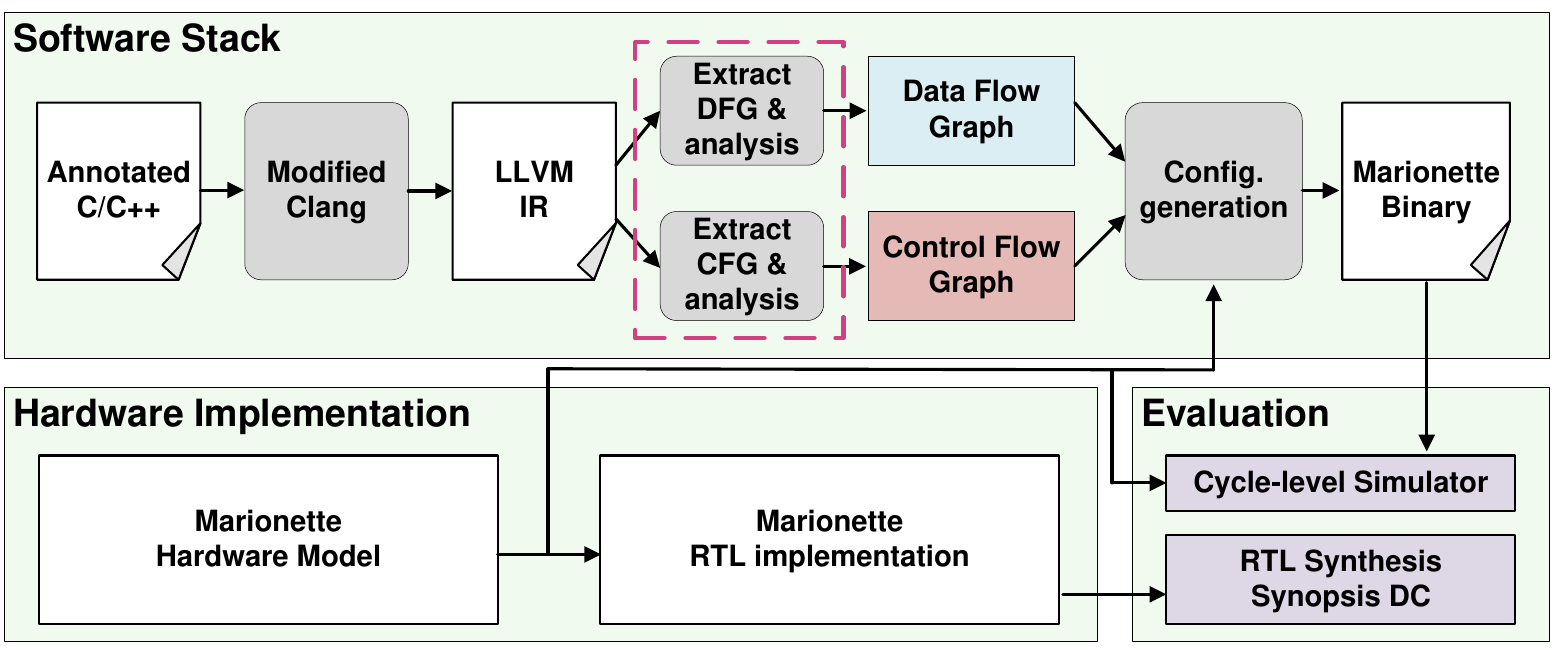}
  \caption{Marionette’s software stack.}
  \label{fig:fig6_1}
\vspace{-15pt}
\end{figure}

\input{MICRO/APtable}

\subsection{Programming and Compilation} \label{subsec:4.4}

%\input{ISA}

%Table \ref{table:tab4} shows the implemented ISA. In addition to the basic operations of the spatial architecture data flow plane, we have added the instruction segment of the control flow plane. It is worth noting that the control input field of ISA are prepared for next configuration. 

The process of programming Marionette entails several tasks: (1) Annotating the branches and loop statements of the algorithm with \#pragma tags, (2) Extracting and analyzing the control data flow graph (CDFG), and (3) Mapping the control flow and data flow portions to the corresponding planes of Marionette, while imposing constraints on memory access and communication. To elucidate the mapping of the program onto Marionette, we employ an example from Figure \ref{fig:fig4_6}. Figure \ref{fig:fig4_6} (a) depicts the original kernel's C code, while Figure \ref{fig:fig4_6} (b) illustrates the extracted CDFG. Subsequently, Figure \ref{fig:fig4_6} (c) showcases the specific mapping onto Marionette's control flow and data flow planes, using the Marionette scheduling algorithm explained in Figure \ref{fig:fig4_5}. The CDFG for Branch Divergence may yield a tree-like structure, and the same level BBs are mapped to the same PE lane to the extent possible. Furthermore, Imperfect Loops are partitioned into distinct mappings.

%% file: MICRO/APtable.tex
\begin{table}[b] 
%\vspace{-15pt}
\setlength{\abovecaptionskip}{0pt}
    
    \renewcommand{\arraystretch}{1.25}
    \centering
    \caption{ \label{table:tab5}Area and power breakdown (28nm)}
    \resizebox{0.9\linewidth}{!}{%
    \begin{tabular}{clcc} 
    \toprule[1.25pt]
    \textbf{Items}                                                                      & \textbf{Component  }                                                                     & \textbf{Area ($mm^2$)} & \textbf{Power ($mW$)} \vspace{2pt}\\ 
    \hline
    
    \textbf{PE}                                                       \rule{0pt}{12pt} & PEs (12 ordinary)                                       & 0.059     & 48.99      \vspace{-2pt}\\
    (60.11\%)\                                                                         & PEs (4 with nonlinear fitting)                                                    & 0.032     & 22.02       \vspace{2pt}\\ 
    \hline
    \textbf{Network}                                                  \rule{0pt}{12pt} & Data Network                                                                    & 0.0063    & 40.80      \vspace{-2pt}\\
    (5.60\%)                                                                           & Control Network                                                                 & 0.0022    & 13.89      \vspace{2pt} \\ 
    \hline
    \multirow{3}{*}{\begin{tabular}[c]{@{}c@{}}\textbf{Memory}\vspace{-2pt}\\(25.58\%)\end{tabular}} 
                                                                      \rule{0pt}{12pt} & Data Scratchpad (16KB)                                                           & 0.033     & 5.07      \vspace{-2pt}\\
                                                                                       & Memory Access Interconnect                                                      & 0.003     & 14.24      \vspace{-2pt}\\
                                                                                       & Control FIFOs                                                                   & 0.001     & 0.56       \vspace{2pt} \\ 
    \hline
    \begin{tabular}[c]{@{}c@{}} \rule{0pt}{12pt} \textbf{Control} \vspace{-2pt} \\(8.71\%)\vspace{2pt} \end{tabular}                  
    & \begin{tabular}[c]{@{}l@{}} \rule{0pt}{12pt}Controller \vspace{-2pt}  \\Instruction Scratchpad (2KB) \vspace{2pt}\end{tabular}                                    & 0.013     & 6.52       \\ 
    \hline
    \textit{\textbf{Total}}                                           \rule{0pt}{12pt} & Marionette                                                     & 0.151     & 152.09    \\
    \bottomrule[1.25pt]
    \end{tabular}
    }
%\vspace{-10pt}
\end{table}

%% file: MICRO/6_Implementation.tex
\section{Implementation}\label{sec:6}

Here we discuss the implementation of the hardware, software stack and simulator used in the evaluation. Figure \ref{fig:fig6_1} shows an overview.

\textbf{Hardware:}
Our Marionette design is parameterizable (e.g., PE array size, FU type, port widths, memory size, etc.) and yields an architectural description shared with the software stack and simulator. Table \ref{table:tab5} shows the hardware parameters. We synthesized a prototype of the Marionette at 500MHz using the 28nm technology library.

\textbf{Software Stack:}
We use the annotated source code to generate an LLVM intermediate representation (IR), which represents low-level operations on data flow and control flow. An automatic tool examines LLVM's IR and generates several DFGs and one CFG based on the PE data flow plane and control flow plane capabilities, respectively. The final bitstream generation step converts CFG and DFG into configuration bitstreams according to the hardware model.

\textbf{Simulator:}
We have developed a cycle-level accurate simulator. It uses the binary configuration file output by the compiler to verify the functional correctness of the Marionette and to evaluate the performance.

%% file: MICRO/7_Evaluation_Methodology.tex
\section{Evaluation Methodology}\label{sec:7}

\begin{table}[b] \footnotesize
%\vspace{-15pt}
\setlength{\abovecaptionskip}{0pt}
        \centering
        \caption{\label{table:tab6}Evaluation benchmarks (All data types are 32-bit).}
        
        \resizebox{0.90\linewidth}{!}{%
        \begin{tabular}{lr}
        \toprule[1pt]
     
            \textbf{Benchmark}            & \textbf{Data Sizes}   \\
        \hline
        \rule{0pt}{8pt}Merge Sort          & 1024     \\
        FFT       & 1024 points      \\
        Viterbi   & 64 stages; 140 obs; 64 tokens        \\
        NW & 128×128     \\
        Hough Transform          & 120×180       \\
        CRC      & 64 bytes   \\
        ADPCM Encode   & 2000 bytes         \\
        SC Decode & 2048 channels       \\
        LDPC Decode   & 20 iters; 128 code length         \\
        GEMM & 64×64     \\
        Conv-1d &  16384  \\
        Sigmoid &  2048    \\
        Gray Processing &  16384\\
        \bottomrule[1pt]
        \end{tabular}
        }
%\vspace{-15pt}
\end{table}

\subsection{Comparison Methodology} \label{subsec:7.1}

We built a cycle-level accurate simulator with the option to implement innovation points to compare the performance gains obtained by each innovation point separately. The simulator optimistically offers high memory access flexibility.

First, we evaluate the performance of Marionette PE (including Proactive PE Configuration) compared to von Neumann PE and dataflow PE. In order to verify the optimization effect on Branch Divergence, for a fair comparison, we do not consider the dedicated control network and Agile PE Assignment. And we unify the data network.

Second, we evaluate the peer-to-peer control network. Besides, we conduct a DC synthesis experiment on the control network delay under different frequencies and network stages. Moreover, we compare the network normalized area with the state-of-the-art architecture.

In addition, we evaluate the Agile PE Assignment to verify the optimization effect on Imperfect Loop. Furthermore, we compute the utilization improvement of the PE that originally executed the outer BB and pipeline utilization, and analyze the relationship between the result of the peer-to-peer control network and the Agile PE Assignment.

Finally, we built the performance models of Softbrain \cite{2017stream}, TIA \cite{TIA}, REVEL \cite{2020REVEL} (15 systolic PEs, 1 tagged-dataflow PE), Riptide \cite{gobieski2022riptide} (16 fully functional PEs and 25 control flow operators inside network) and Marionette with the simulator and normalized the computing fabric to the same size to compare the performance.

%%Conv-1d (CO), Sigmoid (SI) and Gray Processing (GP) are non-complex-CF 这里面的and前面可能要加逗号

\subsection{Benchmark} \label{subsec:7.2} 

We selected a wide range of 13 benchmarks to evaluate our work. FFT, NW, Viterbi (VI), Merge-Sort (MS) and GEMM are from Machsuite \cite{2014machsuite}. ADPCM and CRC are from Mibench \cite{2001mibench}. Hough Transform (HT) is from HosNa Suite \cite{2021hosna}. We also selected LDPC Decode (LDPC) \cite{richardson2008modern} and SC Decode (SCD) \cite{arikan2009channel}. Some of the control flow characteristics of these benchmarks have been qualitatively described in Table \ref{table:tab1}. Conv-1d (CO), Sigmoid (SI) and Gray Processing (GP) are simple single-layer loop applications, which are prepared as a fair comparison. Table \ref{table:tab6} shows the data sizes.

%% file: MICRO/8_Evaluation.tex
\section{Evaluation}\label{sec:8}

We evaluate the performance improvement of features in Marionette. In addition, we conducted scalability experiments for the control network and compared the network with other work. Finally, we compare Marionette with state-of-the-art architectures and show that Marionette performs better on intensive control flow applications.

\begin{figure}[t]
%\vspace{-15pt}
\setlength{\abovecaptionskip}{0pt}
  \centering
  \includegraphics[width=1\linewidth]{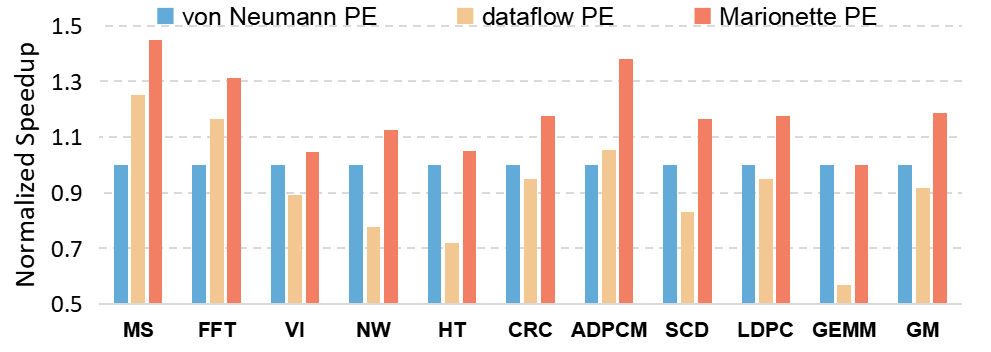}
  \caption{Normalized speedup comparison among von Neumann PE, dataflow PE and Marionette PE (including Proactive PE Configuration): Marionette PE gets a speedup of 1.18× and 1.33× of the von Neumann PE and dataflow PE.}
  \label{fig:exp1}
%\vspace{-15pt}
\end{figure}

\subsection{Advantages of Proactive PE Configuration}\label{subsec:8.1}

Figure \ref{fig:exp1} shows the speedup of our Marionette PE with Proactive PE Configuration compared to von Neumann PE and dataflow PE. The results show that the Marionette PE outperforms von Neumann PE by geomean 1.18× and up to 1.45× (Merge Sort). Moreover, it outperforms dataflow PE by geomean 1.33× and up to 1.76× (GEMM). Further, we separately count the proportion of operators under the branch. The ratio can expose the utilization waste caused by the static mapping of von Neumann execution model. Merge Sort has the highest branch subsequent PE ratio due to the Proactive Configuration saving the most PE resources. Due to the pipeline II, the data flow PE still has poor performance even if it has some flexibility.

\subsection{Advantages of Control Network}\label{subsec:8.2}

\begin{figure}[t]
%\vspace{-15pt}
\setlength{\abovecaptionskip}{3pt}
  \centering
  \includegraphics[width=1\linewidth]{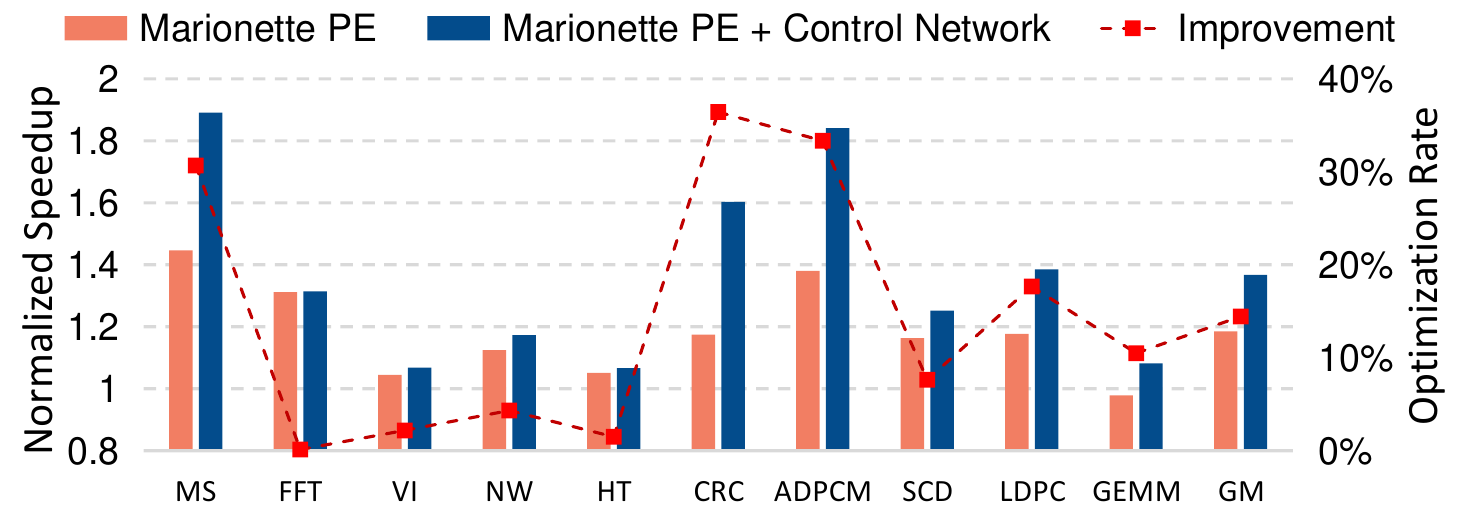}
  \caption{Normalized speedup by Control Network. A peer-to-peer control network contributes a 1.14× performance improvement.}
  \label{fig:exp3}
\vspace{-10pt}
\end{figure}

\begin{table}[b] \LARGE
%\vspace{-15pt}
\setlength{\abovecaptionskip}{0pt}
    \renewcommand{\arraystretch}{1.15}
    \centering
    \caption{\label{table:tab7}Comparison of network area ($mm^2$) in state-of-the-art architectures (normalized to 28nm, 32-bit, $4\times4$ PE array.)}
    
    \resizebox{1\linewidth}{!}{%
    \begin{tabular}{lcccccc}
    \toprule[2pt]
 
        \textbf{architectures}            & \textbf{Softbrain} &\textbf{REVEL}  & \textbf{DySER}  & \textbf{Plasticine} & \textbf{SPU}    & \textbf{Marionette}             \\ 
    \hline
    PE Area~           & 0.0041     & 0.022  & 0.058  & 0.161     & 0.050  & 0.0908           \\
    Network Area       & 0.0130     & 0.028  & 0.052  & 0.294      & 0.045  & \textbf{0.0118}  \\
    Computing Fabric   & 0.0171     & 0.050  & 0.110  & 0.455      & 0.094  & 0.1026           \\
    Network Area Ratio & 75.8\%    & 55.4\% & 47.2\% & 64.6\%     & 47.3\% & \textbf{11.5\%}   \\
    \bottomrule[2pt]
    \end{tabular}
    }
    %\vspace{-10pt}
\end{table}

As shown in Figure \ref{fig:exp3}, the peer-to-peer control network leads to a shorter transfer delay of the control flow. It outperforms on average 1.14× and up to 1.36× (CRC). CRC, ADPCM, and Merge Sort are only partially pipelined. Hence, the overhead of the control flow transfer is high, and the speedup is apparent.

\begin{figure}[t]

\setlength{\abovecaptionskip}{3pt}
  \centering
  \includegraphics[width=1\linewidth]{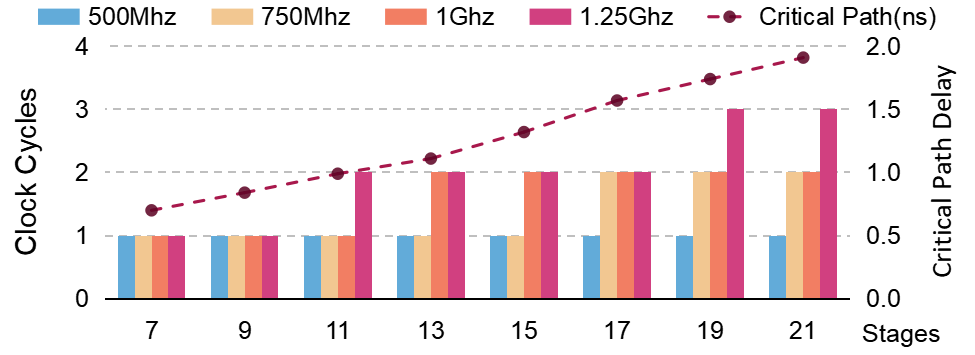}
  \caption{The relationship among network stages, network delay and critical path delay. Control network provides optimal scalability at high frequencies. }
  \label{fig:exp7}
%\vspace{-15pt}
\end{figure}

%% respectively measure the network   respectively后面可能要加一个逗号

After adding the control network, we compare the network area overhead with the state-of-the-art architectures. As shown in Table \ref{table:tab7}, considering a fair comparison, we normalize the computing fabric (Plasticine uses a 3-lane 4-stage PCU and a 4-stage SRAM-free PMU), respectively measure the network (including data network, memory network and control network) area and the ratio of the network to computing fabric. Our network area is only 0.0118$mm^2$, which is 11.5\% of the computing fabric. While each architecture may have a unique functional design with varying PE and network functions and sizes, we can infer from our experiment results that a peer-to-peer control flow network can alleviate the burden of utilizing other network structures for transmitting control flow. This, in turn, reduces the interconnection complexity of the original network and minimizes overhead.

We also evaluate the scalability of the control network by synthesizing different stages of the control network under various time constraints. Figure \ref{fig:exp7} shows the result. Higher frequency and large-scale fabric will increase network latency. However, we believe the low increase in network latency is acceptable because the data flow has more severe constraints than the control flow.

\subsection{Advantages of Agile PE Assignment}\label{subsec:8.3}

\begin{figure}[t]
%\vspace{-15pt}
\setlength{\abovecaptionskip}{3pt}
  \centering
  \includegraphics[width=1\linewidth]{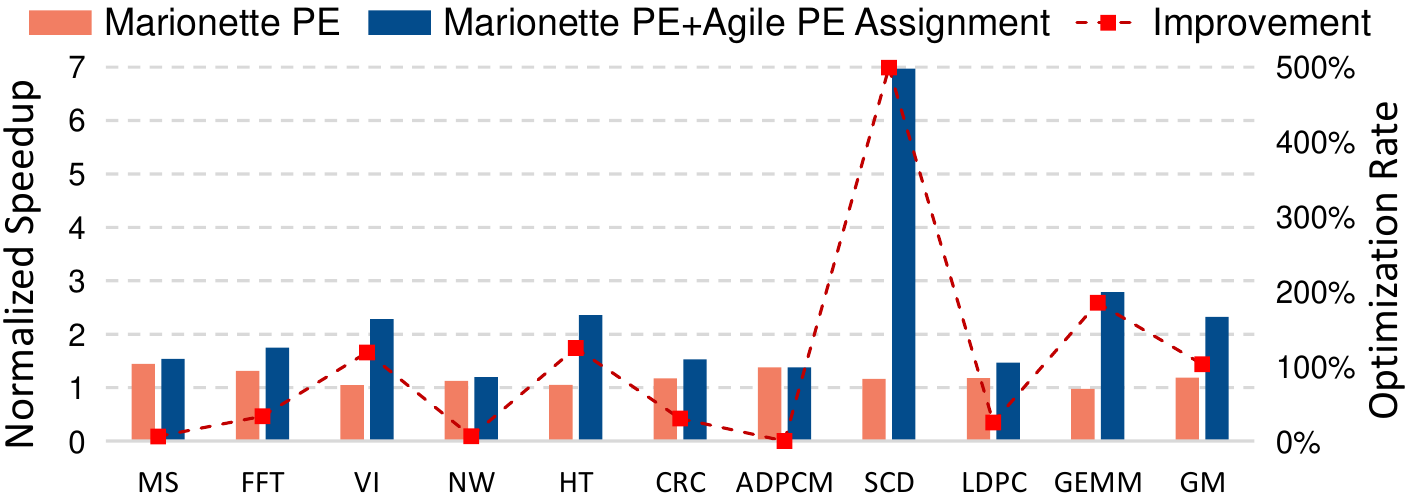}
  \caption{Normalized speedup by Agile PE Assignment. Agile PE Assignment contributes a 2.03x performance improvement. }
  \label{fig:exp3_2}
%\vspace{-10pt}
\end{figure}

\begin{figure}[t]
\setlength{\abovecaptionskip}{0pt}
  \centering
  \includegraphics[width=1\linewidth]{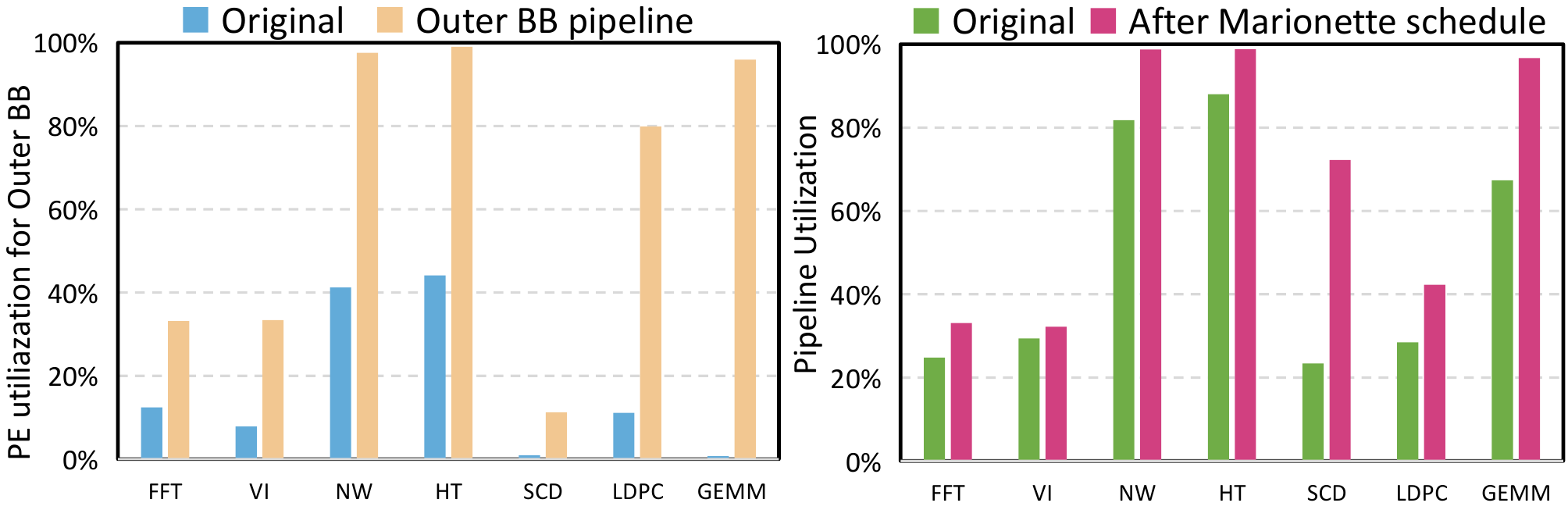}
  \caption{Effects of Agile PE Assignment. outer-BB PE utilization achieves 21.57× optimization, while pipeline utilization achieves 1.54× optimization.}
  \label{fig:exp5}
\vspace{-15pt}
\end{figure}

As shown in Figure \ref{fig:exp3_2}, Agile PE Assignment significantly improves the performance, which achieves an average speedup of 2.03× and up to 5.99×.

We also further analyze the improvement of Agile PE Assignment, respectively from outer-BB PE utilization and pipeline utilization in a fine-grained manner, in Figure\ref{fig:exp5}. We only selected the multi-layer nested loop benchmark where the innermost loop can form a pipeline.

"Outer-BB PE utilization" pertains to those PEs initially assigned to solely execute the outer loop BB. By dynamically assignment, they can either join the construction of the outer loop pipeline or reconfigure them as inner loop pipelines, leading to an average improvement of 21.57×. Among them, GEMM forms a dense spatial pipeline structure and obtains a utilization rate of 134×.

The measure of pipeline utilization is determined by the proportion of pipeline initiations to the overall number of executions. This ratio provides an indication of the pipeline's level of idleness. Overall, Agile PE Assignment has achieved an average of 1.54× optimization in the pipeline utilization in different benchmarks. Hough Transform, NW, SC Decode and GEMM are suitable because outer BBs can generate more control flow. FFT and Viterbi have a data-dependent pipeline II and limit the practical pipeline to 33\% (II=2). In general, the improvement is limited by the loop structure and the limitations of data dependencies between loops (LDPC).

%%between control flow  -> between the control flow 加一个冠词

\begin{figure}[t]
%\vspace{-15pt}
\setlength{\abovecaptionskip}{0pt}
  \centering
  \includegraphics[width=1\linewidth]{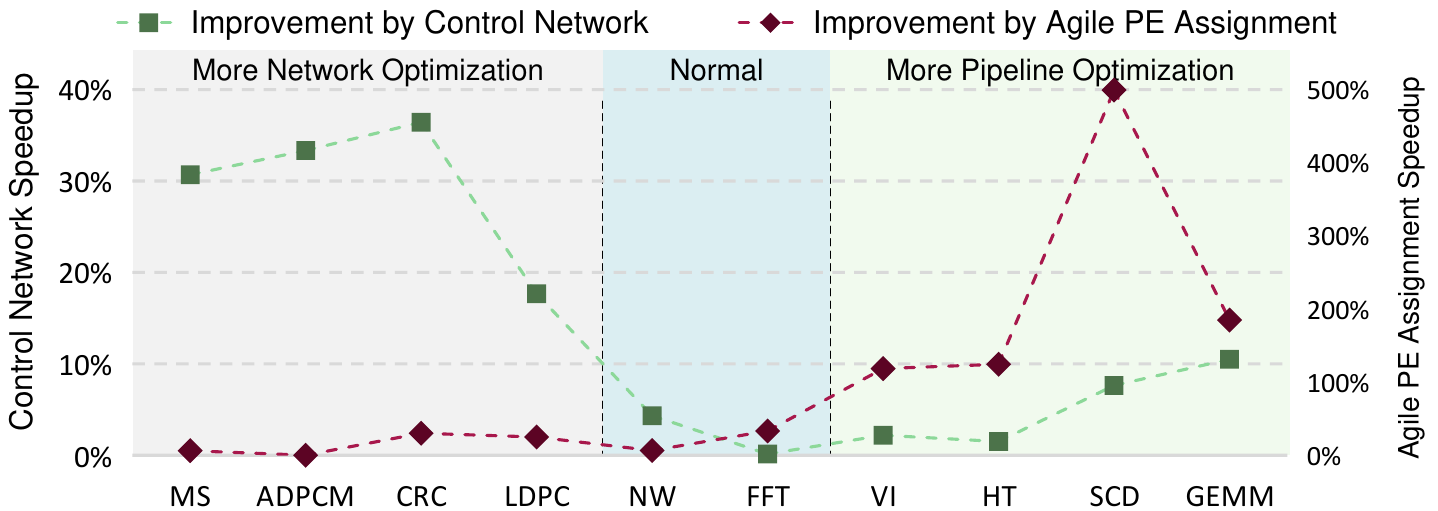}
  \caption{Speedup comparison between control network and Agile PE Assignment. The proportion of benchmarks' control flow that can be hidden distinguishes the acceleration effect of the two features.}
  \label{fig:exp4}
\vspace{-15pt}
\end{figure}

\textbf{Speedup comparison between control network and Agile PE Assignment:} An exciting balance of speedup between the control network and Agile PE Assignment is shown in Figure \ref{fig:exp4}. CRC, ADPCM, Merge Sort, and LDPC cannot be well pipelined. Therefore, Agile PE Assignment cannot create a significant acceleration, but the acceleration of the control network is noticeable. For Viterbi, Hough Transform, SC Decode and GEMM, the control flow is comparatively regular, so the Agile PE Assignment of the control flow is evident. While the proportion of control flow in the critical path is diminished, and the acceleration effect of the control network declines.

\subsection{Marionette Outperforms State-of-the-art architectures}\label{subsec:8.4}

\begin{figure}[t]
\setlength{\abovecaptionskip}{0pt}
  \centering
  \includegraphics[width=1\linewidth]{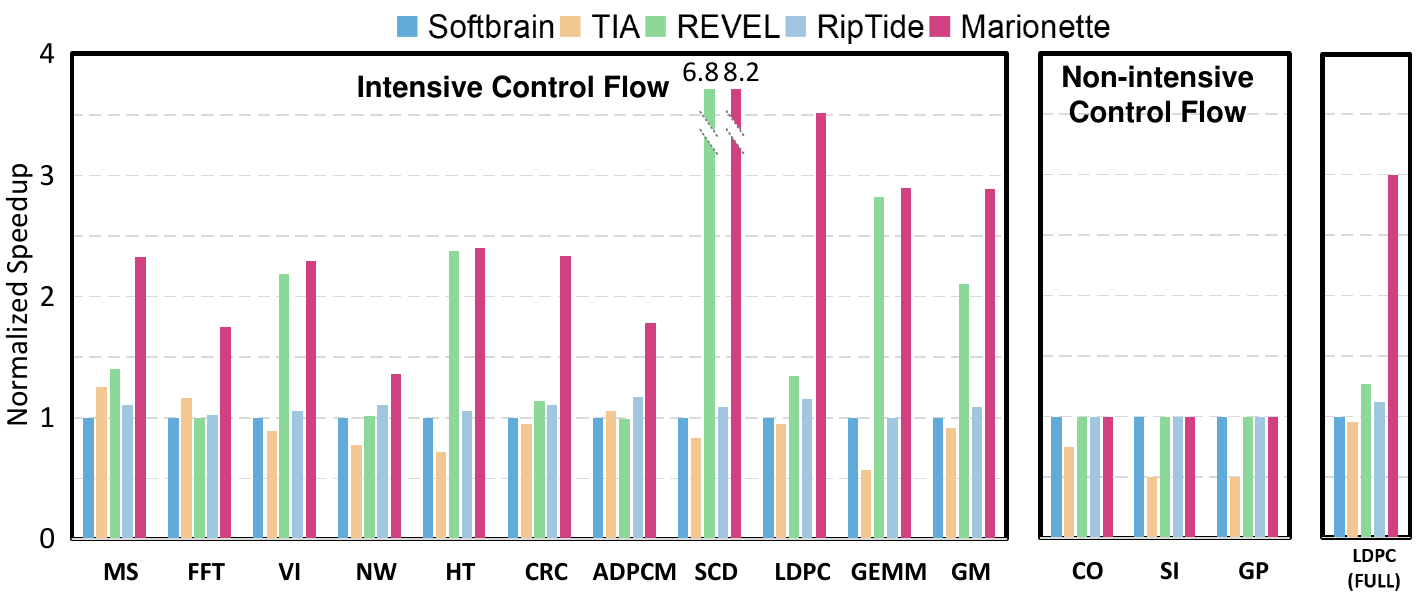}
  \caption{Normalized speedup comparison with state-of-the-art architectures. Marionette does not deteriorate the performance of non-intensive control flow applications. Marionette outperforms Softbrain, TIA, REVEL, and RipTide by geomean 2.88×, 3.38×, 1.55×, and 2.66× on intensive control flow benchmarks. For the full LDPC application containing both intensive control flow and non-intensive control flow kernels, Marionette outperforms Softbrain, TIA, REVEL, and RipTide by 3.01×, 3.13×, 2.36× and 2.68×} 
  \label{fig:exp6}
\vspace{-15pt}
\end{figure}

%% For non-complex-CF benchmarks, all architectures have similar performance xxx 这个句子太长了
%% There is a very small   very small改成 minimal或者 tiny
%%However, in the case where  改成However when
%%Hough Transform, SC Decode and GEMM  and前面加逗号

We compare Marionette with other architectures. The results are shown in Figure \ref{fig:exp6}. For non-intensive control flow benchmarks, all architectures have similar performance except for TIA which has a longer pipeline II (dataflow PE). Single BB inside loop structure is particularly suitable for constructing balanced pipelines. The innovative features of the Marionette do not deteriorate performance for non-intensive control flow applications. For intensive control flow benchmarks, TIA and Softbrain have similar performance. On average, the Marionette speedup is 2.88× that of Softbrain, 3.38× that of TIA, 1.55× that of REVEL, and 2.66× that of RipTide. In some benchmarks such as Viterbi, Hough Transform, SC Decode and GEMM, the REVEL execution model is comparable to the Agile PE Assignment, so the speedup is better. However, Agile PE Assignment has apparent advantages because it is flexible enough.

%% file: MICRO/9_Related_Work.tex
\section{Related Work}\label{sec:9}

\textbf{Spatial architecture taxonomy by execution model:} We divide PEs of SA and reconfigurable spatial architectures into von Neumann PE and dataflow PE according to the execution mode. Table \ref{table:tab8} lists some of them. 
Von Neumann PE only passively executes the configuration according to the instruction sequence. Some SAs \cite{Xputer,piperench,ASH,2007RICA,2009PPA,2009TCPA,2011dyser,2012dyser,2013REMUS,2014FPCA,2015dynaspam,2015nda,2016HRLl,2017HReA,2017stream,2017WaveDPU,2018PX-CGRA,2018i-DPsCGRA,2019spu,MP-CGRA,2022drips,gobieski2022riptide} are configured by a unified controller (main processor or co-processor), and some use counters \cite{2016DORA,2017plasticine} or finite state machines \cite{PADDI,DRP} to control the order of instructions. They both have controllers that construct sequential instruction flows. To satisfy some dynamic properties, there are some von Neumann PEs request configuration from the processor \cite{2006tartan,2007TFLEX,2013T3,2016HARTMP}.
Dataflow PE means the reconfigurable PE determines the execution state according to the input data to select the configuration execution\cite{PADDI2,morphosys,2003wavescalar,2007wavescalar,pact-xpp,2004trips,TIA,2014sSGMF,2018dMT-CGRA}. Its essence is the out-of-order execution of instructions.
The marionette PE is innovative from the von Neumann PE and the data flow PE, and decouples the configuration process through the control plane to achieve timely and Proactive Configuration, which does not exist before.

\textbf{Dedicated Control Network Design:} Some SAs add control bits to tag data with additional functions, such as SPU \cite{2019spu}, etc. The control signal is coupled with the data network, which cannot satisfy control flow flexibility. The DRIPS control network \cite{2022drips} is essentially a config network-on-chip (NoC), not a control network for our control flow characteristics. The RipTide \cite{gobieski2022riptide} moves most control operations to the network, but the transferring is slow and inflexible, and the data and control information in the network are still coupled. Overall, Marionette has the first independent control network designed for control flow from the control plane.

%%by the controller to obtain the current pipeline state and resend xxx   改成动名词 obtaining , resending
%%In our execution model, the pipeline execution resources are dynamically balanced by the control flow.
%%把前面的状语移到最后
%%The pipeline execution resources are dynamically balanced by the control flow in our execution model.

\textbf{Spatial pipelines on multiple BBs:}  Most SAs support spatial pipelines of innermost loops. Some SAs support spatial pipelines of different BBs, but limit their execution resources. FIFER \cite{2021fifer} restricts different BBs to pipeline on different computing fabrics. REVEL \cite{2020REVEL} restricts the innermost loop to pipeline on systolic computing fabric and the outer loop BBs to pipeline on only a few dataflow PEs. The mismatch between the number of pipeline operators and the fixed execution resources can lead to PE underutilization and performance bottlenecks. In our execution model, the BB pipelines are dynamically balanced by the control flow. The mechanism of the DRIPS \cite{2022drips} balancing of pipelines is passive by setting a fixed time window by the controller to obtain the current pipeline state and resend the configuration by the config NoC. The passive centralized balancing approach will lead to pipeline pauses. Our dynamic balancing spatial pipeline is active and decentralized, thus having better dynamic balancing pipeline results.

%Plasticine \cite{2017plasticine} has "nested parallelism" that couples data paths to nested parallelism patterns.

%% file: MICRO/10_Conclusion.tex
\section{Conclusion}\label{sec:10}

This work describes Marionette, a spatial architecture with a decoupled, explicit-designed control flow plane and three corresponding innovative features. We developed full stack of Marionette (ISA, compiler, simulator, RTL) and demonstrate that in a variety of challenging control-intensive applications, compared to state-of-the-art spatial architectures, Marionette outperforms Softbrain, TIA, REVEL, and RipTide by geomean 2.88×, 3.38×, 1.55×, and 2.66×.

%% file: MICRO/11_Acknowledgments.tex
\begin{acks}
This work was supported in part by NSFC Grant 62125403; in part by the Science and Technology Innovation 2030 - New Generation of AI Project under Grant 2022ZD0115200; in part by Beijing Municipal Science and Technology Project Grant Z221100007722023; in part by the National Key Research and Development Program under Grant 2021ZD0114400; in part by Beijing National Research Center for Information Science and Technology; in part by the Beijing Advanced Innovation Center for Integrated Circuits; and in part by Tsinghua University-China Mobile Communications Group Co.,Ltd. Joint Institute.
\end{acks}